\documentstyle[12pt]{article}

\begin{document}
\renewcommand{\theequation}
{\thesection.\arabic{equation}}
\thispagestyle{empty}
\begin{center}
{\LARGE {\bf A Yang--Mills Theory in Loop Space and \\
 \quad \\
Generalized Chapline--Manton Coupling}} \\

\vspace*{20mm}

\renewcommand{\thefootnote}{\fnsymbol{footnote}}

{\Large Tadahito NAKAJIMA 
\footnote[1]{E-mail: nakajima@phys.ge.cst.nihon-u.ac.jp}} \\
\vspace*{20mm}

{\large {\it Physics Laboratory, College of Science and Technology, 
Nihon University}} \\
{\large {\it Funabashi, 274-8501, Japan}} \\ 

\vspace*{20mm}

{\bf Abstract} \\

\end{center}

We consider a Yang-Mills theory in loop space with an affine Lie gauge group. 
The Chapline-Manton coupling, the coupling between Yang-Mills fields and 
an abelian antisymmetric tensor field of second rank via the Chern-Simons 
term, is systematically derived within the framework of the Yang-Mills theory. 
The generalized Chapline-Manton couplings, the couplings among non-abelian 
tensor fields of second rank, Yang-Mills fields, and an abelian 
tensor field of third rank, are also derived by applying the 
non-linear realization method to the Yang-Mills theory. These couplings 
are accompanied by {\it BF}-like terms.

\vspace*{15mm}

\clearpage

\section{Introduction}
\setcounter{page}{1}
\setcounter{equation}{0}

Gauge symmetries have been established as a guiding principle that determines 
couplings among local fields. The fundamental interactions, 
electromagnetic, weak and strong interactions, are intermediated by gauge 
fields, and the couplings between the gauge fields and matter fields and 
the self-coupling of the gauge fields are determined by the 
gauge symmetries. In addition, the gauge symmetries lead to 
the gravitational interaction. Indeed, the gravitational interaction can be 
formulated within the framework of gauge theory based on a non-compact gauge 
group.

In supergravity theories, on the other hand, there is a non-trivial 
interaction which cannot be derived from gauge symmetries alone. In 
${\cal N}\!=\!1$ supergravity theory 
(with ${\cal N}\!=\!1$ super Yang-Mills theory) in ten 
dimensions, the Lagrangian has a coupling between Yang-Mills fields and 
the abelian antisymmetric tensor field in the supergravity multiplet via the 
Chern-Simons 3-form. A non-trivial coupling was first introduced by 
de Wit  et al. in the abelian case, and generalized to the non-abelian 
case by Chapline and Manton. \cite{CM} Although this coupling is determined 
by the local supersymmetry in the supergravity theory, its derivation 
involves much tedious algebra. \cite{ACM} 
In this system, the antisymmetric tensor field must obey 
a deformed transformation rule including the Yang-Mills fields so as to 
maintain the invariance of the Lagrangian. 
This transformation rule can be determined 
uniquely, but only in a heuristic manner. The transformation rule 
plays an important role in proving the Green-Schwarz anomaly cancellation in 
superstring theory. \cite{AC} The non-trivial coupling between Yang-Mills 
fields and the abelian antisymmetric tensor field is referred to as 
the Chapline-Manton coupling. The derivation of the coupling has also been 
discussed from the viewpoint of the BRST cohomology. \cite{CM2}

Recently, the Chapline-Manton coupling has also been derived on the basis of 
a gauge theory --- a Yang-Mills theory in loop space. 
\cite{EYMT} In this theory, the gauge field on loop space is given by the 
functional field on space-time, including an infinite number of local 
component fields. Yang-Mills fields (local Yang--Mills fields) are a part of 
the local component 
fields of the Yang-Mills fields on loop space, while an abelian antisymmetric 
tensor field of second rank is a part of the local component fields of the 
{\it U}(1) gauge field on loop space. 
The couplings among the local component fields of the Yang-Mills fields are 
determined by the symmetry of the loop gauge group. \cite{ALG} 
These couplings are essentially caused by the non-commutativety of the Lie 
algebra. In order to derive non-trivial couplings among the local component 
fields of the Yang-Mills fields and those of the {\it U}(1) gauge field, 
we need an extension of the loop gauge group. The suitable gauge group is the 
affine Lie group, which is a central extension of the loop group. \cite{ALGN} 
Owing to the effect of the central extension, the extended gauge symmetry  
further leads to a new coupling between the local Yang-Mills fields and 
an abelian antisymmetric tensor field of second rank. This coincides with 
the Chapline-Manton coupling. In addition, the deformed 
transformation rule of the abelian antisymmetric tensor field of second 
rank is also derived from the transformation rule of the {\it U}(1) gauge 
field on loop space. \cite{EYMT}

In addition to the abelian antisymmetric tensor field of second rank, 
for example, abelian (totally) antisymmetric tensor fields of higher rank 
appear in the Ramond-Ramond sector of superstring theories. \cite{RR} 
These fields also contribute to a cancellation 
of the anomalies on D-branes under certain conditions. \cite{ABC} 
It is not difficult to extend the Chapline-Manton coupling for 
abelian antisymmetric tensor fields of higher rank. Indeed, such an 
antisymmetric tensor field can couple with the Chern-Simons 
(2n+1)-form. \cite{ABC} \cite{EFF} 
On the other hand, non-abelian antisymmetric tensor fields appear in 
the so-called {\it BF} term. \cite{BF} The {\it BF} term is 
metric independent and takes the form of the product of a non-abelian 
antisymmetric tensor field $B$ and a field strength $F$ of Yang-Mills 
fields in the non-abelian case. The {\it BF} term is a generalization 
of the Chern-Simons term and is an important ingredient in {\it BF} 
Yang-Mills theories, \cite{BFYM} topologically massive gauge theories, 
\cite{TMGT} and so on. However, the extension of the Chapline-Manton 
coupling for non-abelian antisymmetric tensor fields is not known.

As is mentioned above, the Yang-Mills fields and {\it U}(1) gauge field 
on loop space possess an infinite number of local component fields. 
Non-abelian antisymmetric and symmetric tensor fields of second rank are 
the local component fields of the Yang-Mills fields on loop space, 
\cite{LSYM} while an abelian tensor field of third rank with 
certain symmetric properties is the local component field of the {\it U}(1) 
gauge field on loop space. \cite{LSU(1)T} 
The interactions among these non-abelian tensor fields of second rank and 
local Yang-Mills fields can also be derived using the formalism of a 
non-linear realization developed for the loop gauge group applied to 
Yang-Mills theory in loop space. \cite{LSYM} 
In this paper, we consider the Yang-Mills theory in loop space 
with the affine Lie gauge group and apply a non-linear realization method to 
the Yang-Mills theory. As we see below, the non-trivial 
interactions among the local Yang-Mills fields, non-abelian tensor fields 
of second rank, and an abelian tensor field of third rank can be 
systematically determined within the framework of the Yang-Mills theory. 
These local fields interact via a {\it BF}-like term.

This paper is organized as follows. In Sec. 2, we construct a Yang-Mills 
theory in loop space with the affine Lie gauge group. In Sec. 3, it is shown 
that the local field theory for Yang-Mills fields and an antisymmetric 
tensor field with the Chapline-Manton coupling is naturally derived on the 
basis of this theory. In Sec. 4, we apply a non-linear realization method 
developed for the affine Lie gauge group to the Yang-Mills theory in loop 
space. In Sec. 5, 
referring to the previous sections, we derive a local field theory for the 
non-abelian tensor fields, Yang-Mills fields and abelian tensor fields with 
non-trivial coupling. Section 6 is devoted to a summary and discussion of 
the possibilities for future development.


\section{The Yang-Mills theory in loop space}
\setcounter{equation}{0}

We define a loop space $\Omega M^{D}$ as the set of all loops in 
{\it D}-dimensional Minkowski space $M^{D}$. An arbitrary loop 
$x^{\mu}=x^{\mu}(\sigma)\, 
[\,0 \leq \sigma \leq 2\pi\;, x^{\mu}(0)=x^{\mu}(2\pi)\;]$ in $M^{D}$ is 
represented as a point in 
$\Omega M^{D}$ denoted by coordinates $(x^{\mu\sigma})$ with $x^{\mu\sigma} 
\equiv x^{\mu}(\sigma)$.  \footnotemark ${}^{)}$
\footnotetext{${}^{)}$ In the present paper the indices $\mu$, $\nu$, $\kappa$, $\lambda$, $\xi$ and $\zeta$ take the values 0, 1, 2, $\ldots$, {\it D}-1, 
while the indices $\rho$, $\sigma$, $\chi$, and $\omega$ take continuous 
values from 0 to 2$\pi$.}
%

Let us consider a Yang-Mills theory in the loop space $\Omega M^{D}$. We 
assume that a gauge group is an affine Lie group $\widehat{G}_{k}$ whose 
generators $T_{a}(\sigma)$ satisfy the commutation relation 
\begin{eqnarray}
& & {[\, T_{a}(\rho), T_{b}(\sigma) \,]} 
= if_{ab}{}^{c} T_{c}(\rho) \delta(\rho - \sigma) 
+ ik\kappa_{ab} \delta'(\rho - \sigma) \;
\end{eqnarray}
%
and the hermiticity conditions $T^{\dagger}_{a}(\sigma)=T_{a}(\sigma)$.
Here the $f_{ab}{}^{c}$ are the structure constants of the semisimple Lie 
group $G$ and $\kappa_{ab}$ is the Killing metric of $G$. 
\footnotemark ${}^{)}$
\footnotetext{
${}^{)}$ The indices a, b, c, d, and e take the values 1, 2, 3, 
$\ldots$, dim$G$.}
%
The constant $k$ is called the `central charge' of $\hat{G}_{k}$ and takes an 
arbitrary value in this gauge theory. 
When we set $k=0$, the commutation relation (2.1) results in that for the 
loop group $\widehat{G}_{0}$. 
It is possible to make (non-trivial) central extensions to 
infinite-dimensional algebras. \cite{ALG} \cite{KMA} 
Hereafter, we refer to the Yang-Mills theory with an affine Lie gauge 
group as the `extended Yang-Mills theory' (EYMT). 
It will become clear that the central extension of the gauge group leads to 
non-trivial couplings among non-abelian gauge fields and abelian gauge 
fields.

Let ${\cal A}_{\mu\sigma}[x]$ be a gauge field on $\Omega M^{D}$. Owing to 
the central extension, the commutator given in (2.1) yields central terms 
without $T_{a}(\sigma)$, in addition to a linear combination of 
$T_{a}(\sigma)$. Consequently, the gauge field ${\cal A}_{\mu\sigma}$ needs 
extra terms without $T_{a}(\sigma)$ to allow for the consistency of the 
gauge transformation. We define the gauge field ${\cal A}_{\mu\sigma}$ 
as 
\begin{eqnarray}
{\cal A}_{\mu\sigma}[x] 
={\cal A}^{Y}_{\mu\sigma}[x] + \widetilde{\cal A}^{U}_{\mu\sigma}[x] \;,
\end{eqnarray}
%
where ${\cal A}^{Y}_{\mu\sigma}[x]$ is a Yang-Mills field:
\begin{eqnarray}
{\cal A}^{Y}_{\mu\sigma}[x] 
 = \int^{2\pi}_{0} \frac{d\rho}{2\pi} 
 {\cal A}_{\mu\sigma}{}^{a\rho}[x]T_{a}(\rho)\;.
\end{eqnarray}
%
Here ${\cal A}_{\mu\sigma}{}^{a\rho}$ is a vector fields on $\Omega M^{D}$, 
and $\widetilde{\cal A}^{U}_{\mu\sigma}$ is a {\it U}(1) gauge field on 
$\Omega M^{D}$ without $T_{a}(\sigma)$. As in the ordinary 
Yang-Mills theory, the infinitesimal gauge transformation for 
${\cal A}_{\mu\sigma}$ is given by 
\begin{eqnarray}
& & \delta {\cal A}_{\mu\sigma}[x] = 
\partial_{\mu\sigma} \Lambda[x] 
+ i [\,{\cal A}_{\mu\sigma}[x], \Lambda[x]\,] \;,
\end{eqnarray}
%
where $\partial_{\mu\sigma} \equiv \partial/\partial x^{\mu\sigma}$. 
An infinitesimal scalar function $\Lambda$ on $\Omega M^{D}$ is defined 
by
\begin{eqnarray}
\Lambda[x] = \Lambda^{Y}[x] + \Lambda^{U}[x]\;,
\end{eqnarray}
%
where $\Lambda^{Y}[x]$ is written 
\begin{eqnarray}
\Lambda^{Y}[x] 
 = \int^{2\pi}_{0} \frac{d\sigma}{2\pi} 
 \Lambda^{a\sigma}[x]T_{a}(\sigma) \;,
\end{eqnarray}
%
with scalar functions $\Lambda^{a\sigma}$ on $\Omega M^{D}$, and 
$\Lambda^{U}$ is a scalar function $\Omega M^{D}$ without 
$T_{a}(\sigma)$. Since there is no relation between the gauge 
transformation (2.4) and a reparametrization 
$\sigma \rightarrow \bar{\sigma}(\sigma)$,  $\Lambda^{Y}$ and $\Lambda^{U}$ 
obey the following reparametrization invariant conditions: 
\begin{eqnarray}
& & x'^{\mu}(\sigma)\partial_{\mu\sigma}\Lambda^{Y}[x] 
 = \frac{\partial \Lambda^{a\sigma}[x]}{\partial\sigma} \, T_{a}(\sigma) \;, \\
 \nonumber \\
& & x'^{\mu}(\sigma)\partial_{\mu\sigma}\Lambda^{U}[x] = 0 \;. 
\end{eqnarray}
%
Here the prime denotes differentiation with respect to $\sigma$. 
Substituting (2.2) and (2.5) into (2.4) and considering the commutation 
relation (2.1), we obtain (the infinitesimal) gauge 
transformations for ${\cal A}^{Y}_{\mu\sigma}$ and 
$\widetilde{\cal A}^{U}_{\mu\sigma}$: 
\begin{eqnarray}
& & \delta {\cal A}^{Y}_{\mu\sigma}[x] = 
\partial_{\mu\sigma} \Lambda^{Y}[x] 
+ i [\,{\cal A}^{Y}_{\mu\sigma}[x], \Lambda^{Y}[x]\,]^{Y} \;, \\ 
\nonumber \\
& & \delta \widetilde{\cal A}^{U}_{\mu\sigma}[x] = 
\partial_{\mu\sigma} \Lambda^{U}[x] 
+ i [\,{\cal A}^{Y}_{\mu\sigma}[x], \Lambda^{Y}[x]\,]^{U} \;. 
\end{eqnarray}
%
Here, $[\quad, \quad]^{Y}$ denotes the part of a commutator 
$[\quad, \quad]$ written as a linear combination of $T_{a}(\sigma)$, 
while $[\quad, \quad]^{U}$ denotes the other part including the 
central charge $k$. In deriving these gauge transformations, we have used the 
fact that $\widetilde{\cal A}^{U}_{\mu\sigma}$ and $\Lambda^{U}$ are 
commutative. We note that the Yang-Mills fields 
${\cal A}^{Y}_{\mu\sigma}$ appear in the gauge transformation of the 
{\it U}(1) gauge field ${\widetilde {\cal A}}^{U}_{\mu\sigma}$. The 
transformation (2.10) 
is obviously different from an ordinary {\it U}(1) gauge transformation: 
$\delta {\cal A}^{U}_{\mu\sigma} = \partial_{\mu\sigma} \Lambda^{U}$. The 
second term on the right-hand side of (2.10) is due to the central 
extension. Combining the reparametrization invariant condition (2.8) and 
the ordinary {\it U}(1) gauge transformation, 
we can obtain the condition $x'^{\mu}(\sigma){\cal A}^{U}_{\mu\sigma} = 0$. 
\cite{LSU(1)} In the present case, however, the {\it U}(1) gauge field 
$\widetilde{\cal A}^{U}_{\mu\sigma}$ dose not satisfy the condition 
$x'^{\mu}(\sigma)\widetilde{\cal A}^{U}_{\mu\sigma} = 0$ unless $k=0$.

Next, we consider the (naive) field strength of the gauge field 
${\cal A}_{\mu\sigma}$ as 
\begin{eqnarray}
{\cal F}_{\mu\rho, \nu\sigma} 
&=& 
\partial_{\mu\rho}{\cal A}_{\nu\sigma} 
 - \partial_{\nu\sigma}{\cal A}_{\mu\rho}
 + i [\,{\cal A}_{\mu\rho} , {\cal A}_{\nu\sigma}\,] \nonumber \\
 \nonumber \\
&=&
 {\cal F}^{Y}_{\mu\rho, \nu\sigma} + {\cal F}^{U}_{\mu\rho, \nu\sigma} \;,
\end{eqnarray}
%
%
with
\begin{eqnarray}
& & {\cal F}^{Y}_{\mu\rho, \nu\sigma} 
 \equiv \partial_{\mu\rho}{\cal A}^{Y}_{\nu\sigma} 
 - \partial_{\nu\sigma}{\cal A}^{Y}_{\mu\rho}
 + i [\,{\cal A}^{Y}_{\mu\rho} , {\cal A}^{Y}_{\nu\sigma}\,]^{Y} \;, \\
 \nonumber \\
& & {\cal F}^{U}_{\mu\rho, \nu\sigma} 
 \equiv \partial_{\mu\rho}\widetilde{\cal A}^{U}_{\nu\sigma}
 - \partial_{\nu\sigma}\widetilde{\cal A}^{U}_{\mu\rho}
 + i [\,{\cal A}^{Y}_{\mu\rho} , {\cal A}^{Y}_{\nu\sigma}\,]^{U} \;.
\end{eqnarray}
%
%
Note that (2.13) is different from the ordinary field strength of the 
{\it U(1)} gauge field $\widetilde{\cal A}^{U}_{\mu\sigma}$. 
The (naive) field strength (2.11) obeys the 
ordinary gauge transformation rule: $\delta {\cal F}_{\mu\rho, \nu\sigma} 
= i[\,{\cal F}_{\mu\rho, \nu\sigma} , \Lambda \,]$. Because of the central 
extension, however, we can immediately find that 
${\cal F}^{Y}_{\mu\rho, \nu\sigma}$ obeys the homogeneous gauge-transformation 
rule $\delta {\cal F}^{Y}_{\mu\rho, \nu\sigma} 
= i[\,{\cal F}^{Y}_{\mu\rho, \nu\sigma} , \Lambda^{Y}\,]^{Y}$, while 
${\cal F}^{U}_{\mu\rho, \nu\sigma}$ obeys the inhomogeneous 
gauge-transformation rule $\delta {\cal F}^{U}_{\mu\rho, \nu\sigma} 
= i[\,{\cal F}^{Y}_{\mu\rho, \nu\sigma} , \Lambda^{Y}\,]^{U}$. 
Therefore, (2.11) is not suitable for the field strength under the affine Lie 
gauge group. For this reason, we modify (2.11) as 
\begin{eqnarray}
{\cal H}_{\mu\rho, \nu\sigma} 
 \equiv {\cal F}^{Y}_{\mu\rho, \nu\sigma} 
 + {\cal H}^{U}_{\mu\rho, \nu\sigma} \;,
\end{eqnarray}
%
with
\begin{eqnarray}
& & {\cal H}^{U}_{\mu\rho, \nu\sigma} 
 \equiv {\cal F}^{U}_{\mu\rho, \nu\sigma} 
 + k\int\frac{d\omega}{2\pi}x'^{\lambda}(\omega)
 {\rm Tr}\left[ {\cal A}^{Y}_{\lambda\omega}{\cal F}^{Y}_{\mu\rho, \nu\sigma} 
 \right] \;.
\end{eqnarray}
%
Here $``{\rm Tr}"$ denotes the inner product of two elements of the 
affine Lie algebra $\hat{G}$: 
${\rm Tr}[VW] = \sum_{a, b}\int^{2\pi}_{0} \frac{d\sigma}{2\pi} 
\kappa_{ab}V^{a\sigma}W^{b\sigma}$. We can confirm that 
${\cal H}^{U}_{\mu\rho, \nu\sigma}$ is gauge invariant under the 
reparametrization invariant condition (2.7). Note that the gauge invariance 
is still maintained without the central extension. In order for the 
right-hand side of (2.15) to transform in the same manner as 
${\cal F}^{U}_{\mu\rho, \nu\sigma}$ under the reparametrization, however, 
it is necessary to restrict ${\cal A}_{\mu\sigma}{}^{a\rho}$ in (2.4) to the 
form 
\begin{eqnarray}
& & {\cal A}_{\mu\sigma}{}^{a\rho}[x] 
= \delta(\rho - \sigma){\cal A}_{\mu}{}^{a\rho}[x] \;.
\end{eqnarray}
%
Here ${\cal A}_{\mu}{}^{a\rho}$ are the fields on $\Omega M^{D}$ that behave 
as vector functionals on $M^{D}$.

The action for ${\cal A}_{\mu\sigma}$ is defined as 
\begin{eqnarray}
S_{\rm R} = \frac{1}{V_{R}} \int [dx] \left( {\cal L}^{Y} + {\cal L}^{U} \right)  \exp\left( -\frac{L}{l^{2}}\right),
\end{eqnarray}
%
with $V_{R} \equiv \int \{dx\} \exp(-L/l^{2})$, and the Lagrangians
\begin{eqnarray}
& & {\cal L}^{Y} = \frac{1}{4}N^{Y}
 {\cal G}^{\kappa\rho, \lambda\sigma}{\cal G}^{\mu\chi, \nu\omega}
 {\rm Tr}[\,{\cal F}^{Y}_{\kappa\rho, \mu\chi} \; 
 {\cal F}^{Y}_{\lambda\sigma, \nu\omega} \,] \;, \\
 \nonumber \\
& & {\cal L}^{U} = -\frac{1}{4} N^{U}
 {\cal G}^{\kappa\rho, \lambda\sigma}{\cal G}^{\mu\chi, \nu\omega}
 {\cal H}^{U}_{\kappa\rho, \mu\chi}
 {\cal H}^{U}_{\lambda\sigma, \nu\omega} \;,
\end{eqnarray}
%
where $N^{Y}$ and $N^{U}$ are arbitrary constants.

Here, the measures $[dx]$ and $\{dx\}$ are given by $[dx] \equiv 
\prod^{D-1}_{\mu=0}\prod^{\infty}_{n=-\infty}dx^{\mu n}$ and 
$\{dx\} \equiv \prod^{D-1}_{\mu=0}\prod^{\infty, n \neq 0}_{n=-\infty}
dx^{\mu n}$, where the $x^{\mu n}$ are the coefficients of the 
Fourier expansion $x^{\mu}(\sigma) = 
\sum^{\infty}_{n=-\infty}x^{\mu n}e^{in\sigma}$. These measures are 
invariant under a reparametrization. \cite{LSU(1)} 
The (inverse) metric tensor ${\cal G}^{\mu\rho, \nu\sigma}$ on $\Omega M^{D}$ 
is defined by ${\cal G}^{\mu\rho, \nu\sigma} 
\equiv \eta^{\mu\nu} \delta(\rho - \sigma)$, where $\eta_{\mu\nu}$ is 
the metric tensor on $M^{D}$.  \footnotemark ${}^{)}$
\footnotetext{
${}^{)}$ diag $\eta_{\mu\nu} = (1,-1,-1,\ldots,-1)$ \,.}
%
%
The damping factor $\exp(-L/l^{2})$ with $L \equiv -\int^{2\pi}_{0}
\frac{d\sigma}{2\pi}\,\eta_{\mu\nu}\,x'^{\mu}(\sigma)x'^{\nu}(\sigma)$ is 
inserted into the action so that it becomes well defined, where $l\,(> 0)$ is 
a constant with the dimension of length giving the size of loops.

We would like to focus attention on the fact that there is coupling between 
the Yang-Mills fields ${\cal A}^{Y}_{\mu\sigma}$ and the {\it U}(1) 
gauge field ${\cal A}^{U}_{\mu\sigma}$ in the Lagrangian (2.19). It is 
obvious that the coupling is due to the central extension of the gauge group.

The Lagrangians ${\cal L}^{Y}$ and ${\cal L}^{U}$ are gauge invariant, while 
they are not reparametrization invariant, due to the definition of the inner 
product $``{\rm Tr}"$ and the metric tensor ${\cal G}^{\mu\rho, \nu\sigma}$. 
If necessary, we can indeed define an inner product and metric tensor to 
maintain reparametrization invariance as well as gauge-invariance. \cite{LSYM} 
The metric $\eta^{\mu\nu}\delta(\rho-\sigma)$ and the inner product 
$``{\rm Tr}"$ that we employ can be shown in a concrete calculation to be 
forms of the reparametrization invariant inner 
product and metric tensor in a certain gauge of reparametrization.


%
%
\section{The Chapline-Manton coupling}
\setcounter{equation}{0}

In this section, we derive a local field theory with a coupling between 
local Yang-Mills fields and an abelian antisymmetric tensor field of 
second rank on the basis of the Yang-Mills theory in loop space.  Let us 
consider the simplest solutions of (2.7) consisting of local functions on 
$M^{D}$, 
\begin{eqnarray}
\Lambda^{Y(0)}[x] = \int^{2\pi}_{0} \frac{d\sigma}{2\pi}
 g_{0}\,\Lambda^{a}(x(\sigma))\,T_{a}(\sigma) \;, 
\end{eqnarray}
%
where $\Lambda^{a}$ is an infinitesimal scalar function on $M^{D}$ and $g_{0}$ 
is a constant of dimension $[{\rm length}]^{\frac{D-4}{2}}$. On the other 
hand, the simplest solution of (2.8) consisting of a local function on 
$M^{D}$ is given by
\begin{eqnarray}
\Lambda^{U(0)}[x] = \int^{2\pi}_{0} \frac{d\sigma}{2\pi}
q_{0}x'^{\mu}(\sigma)\,\lambda_{\mu}(x(\sigma)) \;, 
\end{eqnarray}
%
%
where $\lambda_{\mu}$ is a vector function on $M^{D}$ and $q_{0}$ is a 
constant of dimension $[{\rm length}]^{\frac{D-6}{2}}$.

Corresponding to (3.1) and (3.2), we consider the (restricted) Yang-Mills 
field ${\cal A}^{Y}_{\mu\sigma}$ and the {\it U}(1) gauge field 
${\cal A}^{U}_{\mu\sigma}$ written in terms of local fields as 
\begin{eqnarray}
& & {\cal A}^{Y(0)}_{\mu\sigma}[x] 
 = g_{0}\,A_{\mu}{}^{a}(x(\sigma))T_{a}(\sigma) \;, \\
\nonumber  \\
& & \widetilde{\cal A}^{U(0)}_{\mu\sigma}[x]
 = q_{0}\,x'^{\nu}(\sigma)\,
 \{ B_{\mu\nu}(x(\sigma)) + C_{\mu\nu}(x(\sigma)) \} \;, 
\end{eqnarray}
%
respectively, where the $A_{\mu}{}^{a}(x)$ are vector fields on $M^{D}$, and 
the $B_{\mu\nu}(x)$ and $C_{\mu\nu}(x)$ are antisymmetric and symmetric 
tensor fields of second rank on $M^{D}$. Obviously, the right-hand sides of 
(3.3) and (3.4) transform in the same manner as the left-hand sides of (3.3) 
and (3.4) under the reparametrization. Substituting (3.1) and (3.3) into (2.9), we obtain the transformation rule of Yang-Mills fields for $A_{\mu}{}^{a}$ as 
\begin{eqnarray}
\delta A_{\mu}{}^{a}(x) = D_{\mu} \Lambda^{a}(x) 
 \equiv \partial_{\mu}\Lambda^{a}(x) 
 - q_{0} A_{\mu}{}^{b}(x)\Lambda^{c}(x)f_{bc}{}^{a} \;,
\end{eqnarray}
%
by virtue of (2.1). On the other hand, substitution of (3.1), (3.2), (3.3) 
and (3.4) into (2.10) yields the transformation rules of $B_{\mu\nu}$ and 
$C_{\mu\nu}$ as 
\begin{eqnarray}
& & \delta B_{\mu\nu}(x) = \partial_{\,[\mu}\lambda_{\nu]}(x) 
 - \tilde{k}_{0}\,A_{[\mu}{}^{a}(x)\partial_{\nu]}\lambda_{a}(x)\;, \\
 \nonumber \\
& & \delta C_{\mu\nu}(x) = 
 - \tilde{k}_{0}\,A_{(\mu}{}^{a}(x)\partial_{\nu)}\lambda_{a}(x)\;,
\footnotemark {}^{)} 
\end{eqnarray}
%
%
where the lowering of the index $a$ has been carried out with $\kappa_{ab}$. 
Here, $\tilde{k}_{0} \equiv kg_{0}/2q_{0}$ is a constant of 
dimension [\,length\,]. Owing to the central extension, the local 
Yang-Mills fields $A_{\mu}{}^{a}$ appear in the transformation rules of 
$B_{\mu\nu}$ and $C_{\mu\nu}$. However, the transformation rules of 
$C_{\mu\nu}$ and $A_{\mu}{}^{a}$ are not independent. 
Indeed, the following symmetric tensor field of second 
rank is invariant under the transformation rules (3.5) and (3.7): 
\footnotetext{${}^{)}$ 
$X_{[\mu}Y_{\nu]} \equiv X_{\mu}Y_{\nu}-X_{\nu}Y_{\mu}, 
\qquad 
X_{(\mu}Y_{\nu)} \equiv X_{\mu}Y_{\nu}+X_{\nu}Y_{\mu}$ \,. } 
%
%
\begin{eqnarray}
\widetilde{C}_{\mu\nu}(x) 
 \equiv C_{\mu\nu}(x) 
 + \frac{1}{2}\tilde{k}_{0}A_{(\mu}{}^{a}(x)A_{\nu),a}(x) \;.
\end{eqnarray}
%

Substituting $C_{\mu\nu} = \widetilde{C}_{\mu\nu}
-\frac{1}{2}\tilde{k}_{0}A_{(\mu}{}^{a}A_{\nu),a}$ into (3.4), 
we obtain $\widetilde{\cal A}^{U(0)}_{\mu\sigma}$ 
written in terms of the local fields 
$A_{\mu}{}^{a}$, $B_{\mu\nu}$ and $\widetilde{C}_{\mu\nu}$. 
As we shall see, the couplings of the local fields are uniquely determined 
in the Yang-Mills theory in loop space. 
On examination of the couplings of these local fields, however, 
we find that the gauge invariant tensor field $\widetilde{C}_{\mu\nu}$ is free 
of $A_{\mu}{}^{a}$ and $B_{\mu\nu}$. \cite{EYMT} 
Since we are interested in the couplings of local fields, we omit 
$\widetilde{C}_{\mu\nu}$ from 
$\widetilde{\cal A}^{U(0)}_{\mu\sigma}$, written in terms of the local fields 
$A_{\mu}{}^{a}$, $B_{\mu\nu}$ and $\widetilde{C}_{\mu\nu}$ for simplicity. 
In other words, we replace $C_{\mu\nu}$ with 
$-\frac{1}{2}\tilde{k}_{0}A_{(\mu}{}^{a}A_{\nu),a}$ in (3.4). 
Then (3.4) is rewritten as 
\begin{eqnarray}
\widetilde{\cal A}^{U(0)}_{\mu\sigma}[x]
 = {\cal A}^{U(0)}_{\mu\sigma}[x] 
 -\frac{1}{2}\tilde{k}_{0}q_{0} \,x'^{\nu}(\sigma)\,
  A_{(\mu}{}^{a}(x(\sigma))A_{\nu),a}(x(\sigma)) 
  \;, 
\end{eqnarray}
%
with
\begin{eqnarray}
{\cal A}^{U(0)}_{\mu\sigma}[x]
 = q_{0}\,x'^{\nu}(\sigma)B_{\mu\nu}(x(\sigma)) \;,
\end{eqnarray}
%
where ${\cal A}^{U(0)}_{\mu\sigma}$ is written in terms of the abelian 
local field only. Since the constant $\tilde{k}_{0}$ is proportional to 
the central charge $k$, 
the {\it U}(1) gauge field $\widetilde{\cal A}^{U(0)}_{\mu\sigma}$ is reduced 
to ${\cal A}^{U(0)}_{\mu\sigma}$ by setting $k=0$. 
Note that $\widetilde{\cal A}^{U(0)}_{\mu\sigma}$ does not satisfy 
the condition $x'^{\mu}(\sigma)\widetilde{\cal A}^{U(0)}_{\mu\sigma}=0$ 
for $k \neq 0$, 
while ${\cal A}^{U(0)}_{\mu\sigma}$ satisfies the 
condition $x'^{\mu}(\sigma){\cal A}^{U(0)}_{\mu\sigma}=0$. 
As we have discussed in Sec. 2, this fact is consistent 
with the fact that the {\it U}(1) gauge field 
$\widetilde{\cal A}^{U}_{\mu\sigma}$ 
does not satisfy the condition 
$x'^{\mu}(\sigma)\widetilde{\cal A}^{U}_{\mu\sigma} = 0$ unless $k=0$.

Next, let us consider the field strength. Substituting (3.3) 
into (2.12), by virtue of (2.1) we obtain 
${{\cal F}}^{Y(0)}_{\mu\rho, \nu\sigma}$ 
written in terms of the field strength of $A_{\mu}{}^{a}$: 
\begin{eqnarray}
{\cal F}^{Y(0)}_{\mu\rho, \nu\sigma}
 = g_{0}\,F_{\mu\nu}{}^{a}(x(\sigma))\,T_{a}(\sigma)\,\delta(\rho-\sigma) \;, 
\end{eqnarray}
%
with
\begin{eqnarray}
F_{\mu\nu}{}^{a}(x) 
 = \partial_{\mu}A_{\nu}{}^{a}(x)-\partial_{\nu}A_{\mu}{}^{a}(x) 
 - A_{\mu}{}^{b}(x)A_{\nu}{}^{c}(x)f_{bc}{}^{a} \;. 
\end{eqnarray}
%
Also the substitution of (3.3), (3.4) and 
(3.11) into (2.15) yields 
\begin{eqnarray}
{\cal H}^{U(0)}_{\mu\rho, \nu\sigma} 
 = q_{0}\,x'^{\lambda}(\sigma) H_{\mu\nu\lambda}(x(\sigma)) 
\end{eqnarray}
%
with 
\begin{eqnarray}
H_{\mu\nu\kappa}(x) 
 = F_{\mu\nu\kappa}(x) + \tilde{k}_{0}\Omega_{\mu\nu\kappa}(x) \;,
\end{eqnarray}
%
and
\begin{eqnarray}
& & F_{\mu\nu\lambda}(x)
 \equiv \partial_{\mu}B_{\nu\lambda}(x)
 + \partial_{\nu}B_{\lambda\mu}(x) + \partial_{\lambda}B_{\mu\nu}(x) \;, \\
 \nonumber \\
& & \Omega_{\mu\nu\lambda}(x)
 = A_{[\mu}{}^{a}(x)\partial_{\nu}A_{\lambda], a}(x) 
 - \frac{g_{0}}{3}A_{[\mu}{}^{a}(x)A_{\nu}{}^{b}(x)A_{\lambda]}{}^{c}(x)
 f_{abc} 
 \;.
\end{eqnarray}
%
Here, $\Omega_{\mu\nu\lambda}$ occurring in $H_{\mu\nu\lambda}$ is a 
Chern-Simons 3-form. Reflecting the fact that 
$\widetilde{{\cal F}}^{U(0)}_{\mu\rho, \nu\sigma}$ is gauge invariant, 
$H_{\mu\nu\lambda}$ becomes also invariant under 
the transformation rules of (3.5) and (3.6).

Finally, let us derive the action $S^{(0)}_{R}$ of the local fields 
$A_{\mu}{}^{a}$ and $B_{\mu\nu}$. Substituting (3.11) and (3.13) into (2.18) 
and (2.19), respectively, and integrating them over $\rho$, $\chi$ 
and $\omega$, we obtain the Lagrangians ${\cal L}^{Y(0)}$ and 
${\cal L}^{U(0)}$ expressed as integrals over $\sigma$: 
\begin{eqnarray}
& & {\cal L}^{Y(0)} = -\frac{1}{4}N^{Y}g_{0}^{2}\delta(0)^{2}
 \int^{2\pi}_{0}\frac{d\sigma}{2\pi}
 F_{\mu\nu, a}(x(\sigma))F^{\mu\nu, a}(x(\sigma)) \;, \\
 \nonumber \\
& & {\cal L}^{U(0)} = -\frac{1}{4}N^{U}q_{0}^{2}\delta(0)
 \int^{2\pi}_{0}\frac{d\sigma}{2\pi}
 x'^{\kappa}(\sigma)x'_{\lambda}(\sigma)
 H_{\kappa\mu\nu}(x(\sigma))H^{\lambda\mu\nu}(x(\sigma)) \;. 
\end{eqnarray}
%
%
We next insert (3.17) and (3.18) into (2.17) and expand the 
functions of $x^{\mu}(\sigma)$, 
$F_{\mu\nu, a}(x(\sigma)) F^{\mu\nu, a}(x(\sigma))$ and 
$H_{\mu\nu\lambda}(x(\sigma)) H^{\mu\nu\lambda}(x(\sigma))$, about 
$x^{\mu 0}$. Then, all the differential coefficients at 
$x^{\mu 0}$ in each Taylor series become total derivatives with respect to 
$x^{\mu 0}$ and vanish under the boundary conditions 
with $|x^{\mu 0}| \rightarrow \infty$. As a result, we obtain an action in 
which the argument $x^{\mu}(\sigma)$ of the functions is replaced with 
$x^{\mu 0}$.

Carrying out the integrations with respect to $x^{\mu n}$ after the 
Wick rotations $x^{\mu n} \rightarrow -ix^{\mu n}\,(n \neq 0)$, 
we obtain the action $S^{(0)}_{{\rm R}}$ as 
\begin{eqnarray}
S_{{\rm R}}^{(0)} = \int d^{D}x \left\{ \,
 - \frac{1}{4} F_{\mu\nu}{}^{a}(x)F^{\mu\nu}{}_{a}(x)
 + \frac{1}{12} H_{\mu\nu\lambda}(x)H^{\mu\nu\lambda}(x) 
 \, \right\} \;.
\end{eqnarray}
%
Here, we set the normalization conditions as $N^{Y}g_{0}^{2}\delta(0)=1$ and 
$3N^{U}q_{0}^{2}l^{2}\delta^{2}(0)/2=1$. Thus, we obtain the action 
describing the system with the coupling between the local Yang-Mills 
fields $A_{\mu}{}^{a}$ and the abelian antisymmetric tensor field $B_{\mu\nu}$ 
via the Chern-Simons 3-form. Setting the structure constants $f_{ab}{}^{c}$ 
to $0$ in (2.1), we can obtain the action (3.19) in the abelian 
version. It describes the system with the coupling between the {\it U}(1) 
gauge field $A_{\mu}$ and the abelian antisymmetric tensor field $B_{\mu\nu}$ 
via the abelian Chern-Simons 3-form. Such couplings have been 
introduced by de Wit et al. in the abelian case and 
Chapline and Manton in the non-abelian case. \cite{CM} \cite{ACM} They assign 
the transformation rule (3.6) to the 
antisymmetric tensor field $B_{\mu\nu}$ in order for (3.14) to be gauge 
invariant. In contrast, we can naturally derive (3.6) in the framework of the 
gauge theory.


%
\section{Application of the non-linear realization to the EYMT}
\setcounter{equation}{0}

In this section, we discuss the local field theories for 
higher rank tensor fields based on the EYMT.

Let us consider the solutions of (2.7) consisting of local functions 
on $M^{D}$: 
\begin{eqnarray}
\Lambda^{Y(p)}[x] = \int^{2\pi}_{0} \frac{d\sigma}{2\pi}
 g_{p}Q^{\mu_{1}}(\sigma)Q^{\mu_{2}}(\sigma) \cdots Q^{\mu_{p}}(\sigma)\,
 \lambda_{\mu_{1}\mu_{2},\ldots,\mu_{p}}{}^{a}(x(\sigma))T_{a}(\sigma) \;. 
\end{eqnarray}
%
Here, $Q_{\mu}(\sigma) \equiv x'^{\mu}(\sigma)/ \sqrt{-x'^{\,2}(\sigma)}$,
\footnotemark ${}^{)}$
\footnotetext{${}^{)}$ $x'^{\,2}(\sigma) \equiv 
x'_{\mu}(\sigma)x'^{\mu}(\sigma)$ \,. }
%
%
where the $\lambda_{\mu_{1}\mu_{2},\ldots,\mu_{p}}{}^{a}(x)$ are infinitesimal 
tensor function of rank $p$ ($p=0,1,2,\ldots,$) on $M^{D}$ and $g_{p}$ 
is a constant of dimension $[{\rm length}]^{\frac{D-4}{2}}$. 
Similarly, the solutions of (2.8) are given by 
\begin{eqnarray}
\Lambda^{U(p)}[x] \!\!&=&\!\! \int^{2\pi}_{0} \frac{d\sigma}{2\pi}
 q_{p}\sqrt{-x'^{\,2}(\sigma)} Q^{\mu_{1}}(\sigma)Q^{\mu_{2}}(\sigma) \cdots 
 Q^{\mu_{p}}(\sigma)Q^{\mu_{p+1}}(\sigma)\, 
 \lambda_{\mu_{1}\mu_{2},\ldots,\mu_{p}\mu_{p+1}}(x(\sigma)) 
\nonumber \\
\!\!&+&\!\! \int^{2\pi}_{0} \frac{d\sigma}{2\pi} 
 e_{p}Q'^{\mu_{1}}(\sigma)Q^{\mu_{2}}(\sigma) \cdots 
 Q^{\mu_{p}}(\sigma)\, \kappa_{\mu_{1}\mu_{2},\ldots,\mu_{p}}(x(\sigma)) \;, 
\end{eqnarray}
%
where $\lambda_{\mu_{1}\mu_{2},\ldots,\mu_{p}\mu_{p+1}}(x)$ and 
$\kappa_{\mu_{1}\mu_{2},\ldots,\mu_{p}}(x)$ are infinitesimal 
tensor functions of rank ($p$ +1) and $p$ on $M^{D}$, 
and $q_{p}$ and $e_{p}$ are constants of dimension 
$[{\rm length}]^{\frac{D-6}{2}}$ and $[{\rm length}]^{\frac{D-4}{2}}$, 
respectively. Setting $p=0$, the infinitesimal functions (4.1) and (4.2) 
correspond to (3.1) and (3.2), respectively.

Any general solution of (2.7) is given as a linear combination of 
$\Lambda^{Y(p)}$, while any general solution of (2.8) is given as a linear 
combination of $\Lambda^{U(p)}$. Explicitly, $\Lambda^{Y}[x]$ and 
$\Lambda^{U}[x]$ can be expressed as $\Lambda^{Y}[x] \equiv 
\sum^{\infty}_{p=0}\Lambda^{Y(p)}[x]$ 
and $\Lambda^{U}[x] \equiv \sum^{\infty}_{p=0}\Lambda^{U(p)}[x]$. 
(The coefficients of the terms in these sums are absorbed into 
$g_{p}$ and $q_{p}$.) 
Consequently, any infinitesimal function $\Lambda$ consisting of local 
functions on $M^{D}$ is given in the form of a linear combination: 
\begin{eqnarray}
\Lambda[x] = \Lambda^{Y}[x] + \Lambda^{U}[x] 
\equiv \sum^{\infty}_{p=0} \Lambda^{(p)}[x] \;, 
\end{eqnarray}
%
with 
\begin{eqnarray}
\Lambda^{(p)}[x] \equiv \Lambda^{Y(p)}[x] + \Lambda^{U(p)}[x] \;.
\end{eqnarray}
%
Corresponding to (4.3), we express the gauge field 
${\cal A}_{\mu\sigma}$ as a linear combination of ${\cal A}^{(p)}_{\mu\sigma}$ 
associated with $\Lambda^{(p)}$ :
\begin{eqnarray}
{\cal A}_{\mu\sigma}[x]
={\cal A}^{Y}_{\mu\sigma}[x] + \widetilde{{\cal A}}^{U}_{\mu\sigma}[x]
\equiv \sum^{\infty}_{p=0} {\cal A}^{(p)}_{\mu\sigma}[x] \;,
\end{eqnarray}
%
with 
\begin{eqnarray}
{\cal A}^{(p)}_{\mu\sigma}[x] \equiv 
{\cal A}^{Y(p)}_{\mu\sigma}[x] + \widetilde{{\cal A}}^{U(p)}_{\mu\sigma}[x] 
 \;,
\end{eqnarray}
%
where ${\cal A}^{Y(p)}_{\mu\sigma}$ are Yang-Mills fields consisting of the 
local tensor fields of rank $p$ and $(p+1)$ on $M^{D}$, and 
$\widetilde{{\cal A}}^{U(p)}_{\mu\sigma}$ are {\it U}(1) gauge fields 
consisting of 
the local tensor fields of rank $p$, $(p+1)$ and $(p+2)$ on $M^{D}$. 
(As an exceptional case, ${\cal A}^{Y(0)}_{\mu\sigma}$ consists of 
local vector fields only, and $\widetilde{{\cal A}}^{U(0)}_{\mu\sigma}[x]$ 
consists of local fields of second rank only.)

Substituting (4.3) and (4.5) into (2.2), and comparing the two sides of the 
resulting equation, we conclude that
\begin{eqnarray}
\delta {\cal A}^{(p)}_{\mu\sigma} 
 = \partial_{\mu\sigma} \Lambda^{(p)} 
 + i \sum^{p}_{k=0} [\, {\cal A}^{(k)}_{\mu\sigma}, \Lambda^{(p-k)} \,] \;.
\end{eqnarray}
%
We note that ${\cal A}^{(0)}_{\mu\sigma}$ obeys the same 
gauge transformation as (2.2), while the ${\cal A}^{(p)}_{\mu\sigma}\;\;
(p=1,2,3,\ldots)$ do not obey the gauge transformation as (2.2). 
Indeed, the gauge transformation of 
${\cal A}^{(p)}_{\mu\sigma}\;\;(p=1,2,3,\ldots)$ depends on other gauge 
fields ${\cal A}^{(k)}_{\mu\sigma}$ $[k\,(<p)=0,1,2,3,\ldots]$.

Next, we consider the (naive) field strength of ${\cal A}^{(p)}_{\mu\sigma}$. 
We substitute (4.5) into (2.11). Then we can 
decompose ${\cal F}_{\mu\rho, \nu\sigma}$ as 
\begin{eqnarray}
{\cal F}_{\mu\rho, \nu\sigma} 
= {\cal F}^{Y}_{\mu\rho, \nu\sigma} 
+ {\cal F}^{U}_{\mu\rho, \nu\sigma}
= \sum^{\infty}_{p=0}{\cal F}^{(p)}_{\mu\rho, \nu\sigma} \;,
\end{eqnarray}
%
with
\begin{eqnarray}
{\cal F}^{(p)}_{\mu\rho, \nu\sigma} \equiv
 \partial_{\mu\rho}{\cal A}^{(p)}_{\nu\sigma}
-\partial_{\nu\sigma}{\cal A}^{(p)}_{\mu\rho}
+ \sum^{p}_{k=0} [\,{\cal A}^{(k)}_{\mu\rho}, 
 {\cal A}^{(p-k)}_{\nu\sigma}\,] \;.
\end{eqnarray}
%
Under the gauge transformation (4.7), the 
${\cal F}^{(p)}_{\mu\rho, \nu\sigma}$ transform as
\begin{eqnarray}
\delta {\cal F}^{(p)}_{\mu\rho, \nu\sigma} 
= i \sum^{p}_{k=0} 
 [\,{\cal F}^{(k)}_{\mu\rho, \nu\sigma}, \Lambda^{(p-k)} \,] \;.
\end{eqnarray}
%

From (4.10), we see that the ${\cal F}^{(p)}_{\mu\rho, \nu\sigma} \,
(p=1,2,3,\ldots)$ do not obey the same transformation rule 
$\delta {\cal F}_{\mu\rho, \nu\sigma} 
= i[\,{\cal F}_{\mu\rho, \nu\sigma} , \Lambda \,]$, 
except ${\cal F}^{(0)}_{\mu\rho, \nu\sigma}$. 
Accordingly, we cannot construct the ``modified" field strengths 
corresponding to (2.14) for ${\cal A}^{(p)}_{\mu\sigma}\;(p=1,2,3,\ldots,)$ 
in the manner discussed in Sec. 2. To begin with, we must 
find the suitable field strengths of ${\cal A}^{(p)}_{\mu\sigma}
\;(p=1,2,3,\ldots,)$ that obey the same transformation rule 
as ${\cal F}_{\mu\rho, \nu\sigma}$. Such field strengths can be 
systematically derived by using a non-linear realization method.

Let us consider the linear subspace $\widehat{\rm {\bf g}}_{k}^{(p)}$ all of 
whose elements have the form 
\begin{eqnarray}
\Xi^{(p)}[x] \equiv \Xi^{Y(p)}[x] + \Xi^{U(p)}[x] \;,
\end{eqnarray}
%
with
\begin{eqnarray}
\Xi^{Y(p)}[x] \!\!&=&\!\! \int^{2\pi}_{0} \frac{d\sigma}{2\pi}
 g_{p}Q^{\mu_{1}}(\sigma)Q^{\mu_{2}}(\sigma) \cdots Q^{\mu_{p}}(\sigma)\,
 \xi_{\mu_{1}\mu_{2},\ldots,\mu_{p}}{}^{a}(x(\sigma))T_{a}(\sigma) \;, \\
\nonumber \\
\Xi^{U(p)}[x] \!\!&=&\!\! \int^{2\pi}_{0} \frac{d\sigma}{2\pi}
 q_{p}\sqrt{-x'^{\,2}(\sigma)} Q^{\mu_{1}}(\sigma)Q^{\mu_{2}}(\sigma) \cdots 
 Q^{\mu_{p}}(\sigma)Q^{\mu_{p+1}}(\sigma)\, 
 \xi_{\mu_{1}\mu_{2},\ldots,\mu_{p}\mu_{p+1}}(x(\sigma)) 
\nonumber \\
\!\!&+&\!\! \int^{2\pi}_{0} \frac{d\sigma}{2\pi} 
 e_{p}Q'^{\mu_{1}}(\sigma)Q^{\mu_{2}}(\sigma) \cdots 
 Q^{\mu_{p}}(\sigma)\, \zeta_{\mu_{1}\mu_{2},\ldots,\mu_{p}}(x(\sigma)) \;.
\end{eqnarray}
%
%
%
Here, $\xi_{\mu_{1}\mu_{2},\ldots,\mu_{p}}{}^{a}(x)$ and 
$\zeta_{\mu_{1}\mu_{2},\ldots,\mu_{p}}(x)$ are arbitrary tensor 
functions of rank $p$ and $\xi_{\mu_{1}\mu_{2},\ldots,\mu_{p}\mu_{p+1}}(x)$ 
is an arbitrary tensor function of rank ($p$ +1) on $M^{D}$. 
Using (2.1), we obtain a commutation relation for every 
$\Xi^{(p)}[x] \in \widehat{\rm {\bf g}}_{k}^{(p)}$ and 
$\Xi^{(q)}[x] \in \widehat{\rm {\bf g}}_{k}^{(q)}$ as
\begin{eqnarray}
[\,\Xi^{(p)},\, \Xi^{(q)}\,] \in \widehat{\rm {\bf g}}_{k}^{(p+q)} \;.
\end{eqnarray}
%
This commutation relation shows that the linear subspace 
$\widehat{\rm {\bf g}}_{k}^{(0)}$ is a Lie algebra, whereas 
$\widehat{\rm {\bf g}}_{k}^{(p)} \;(p=1,2,3,\ldots)$ is not a 
Lie algebra. The direct sum 
$\widehat{\rm {\bf g}}_{k}$ $\equiv \bigoplus^{\infty}_{p=0}$ 
$\widehat{\rm {\bf g}}_{k}^{(p)}$ is obviously a Lie algebra. 
Thus, the linear subspace $\widehat{\rm {\bf g}}_{k}^{(0)}$ forms 
a subalgebra of $\widehat{\rm {\bf g}}_{k}$. 
We consider the Lie groups $\widehat{\rm {\bf G}}_{k}$ 
and $\widehat{\rm {\bf G}}^{(0)}_{k}$ associated with 
$\widehat{\rm {\bf g}}_{k}$ and 
$\widehat{\rm {\bf g}}^{(0)}_{k}$, respectively. Here, both 
$\widehat{\rm {\bf G}}_{k}$ and $\widehat{\rm {\bf G}}^{(0)}_{k}$ 
are Lie subgroups of the affine Lie group $\widehat{G}_{k}$. 
Since 
$\widehat{\rm {\bf G}}^{(0)}_{k}$ is a subgroup of 
$\widehat{\rm {\bf G}}_{k}$, we can consider the coset manifold 
$\widehat{\rm {\bf G}}_{k}$/
$\widehat{\rm {\bf G}}^{(0)}_{k}$. 
We introduce the scalar field 
$\Phi^{(p)}[x]$ on loop space so as to parameterize the coset 
representative of 
$\widehat{\rm {\bf G}}_{k}$/$\widehat{\rm {\bf G}}^{(0)}_{k}$: 
\begin{eqnarray}
{\cal V}[\Phi] = \exp \left( -i\sum^{\infty}_{p=1}\Phi^{(p)}[x] \right) \;. 
\end{eqnarray}
%
Here $\Phi^{(p)}[x]$ is an element of $\widehat{\rm {\bf g}}^{(p)}_{k}$: 
\begin{eqnarray}
\Phi^{(p)}[x] = \Phi^{Y(p)}[x] + \Phi^{U(p)}[x] \;.
\end{eqnarray}
%
The (finite) transformation rule of 
${\cal V}[\Phi] \rightarrow {\cal V}[\bar{\Phi}]$ for 
${\cal X} \in \widehat{\rm {\bf G}}_{k}$ is given by 
\begin{eqnarray}
{\cal V}[\Phi] \rightarrow 
{\cal V}[\bar{\Phi}] = {\cal X} {\cal V}[\Phi] {\cal Y}^{-1}[\Phi, {\cal X}]\;,
\end{eqnarray}
%
where ${\cal X} = \exp(-i\sum^{\infty}_{p=0}\Xi^{(p)})$ and 
${\cal Y} = \exp(-i\Xi^{(0)})$. 
(Here, ${\cal Y}$ depends on ${\cal X}$ and $\Phi$.)

Next, we define the vector field $\widehat{\cal A}_{\mu\sigma}$ by using 
${\cal A}_{\mu\sigma}$ and ${\cal V}[\Phi]$: 
\begin{eqnarray}
\widehat{\cal A}_{\mu\sigma} \equiv 
{\cal V}^{-1}{\cal A}_{\mu\sigma}{\cal V} 
- i{\cal V}^{-1} \partial_{\mu\sigma} {\cal V} \;.
\end{eqnarray}
%
Substituting (4.5) and (4.15) into (4.18), we can express 
$\widehat{\cal A}_{\mu\sigma}$ as a linear combination: 
$\widehat{\cal A}_{\mu\sigma} = 
\sum^{\infty}_{p=0}\widehat{\cal A}^{(p)}_{\mu\sigma}$. The concrete forms of 
$\widehat{\cal A}^{(0)}_{\mu\sigma}$ and $\widehat{\cal A}^{(1)}_{\mu\sigma}$ 
are given by 
\begin{eqnarray}
& & \widehat{\cal A}^{(0)}_{\mu\sigma} = {\cal A}^{(0)}_{\mu\sigma} \;, \\
\nonumber \\
& & \widehat{\cal A}^{(1)}_{\mu\sigma} 
= {\cal A}^{(1)}_{\mu\sigma} - (\partial_{\mu\sigma}\Phi^{(1)} 
+ i[\,{\cal A}^{(0)}_{\mu\sigma}, \Phi^{(1)}\,]) \;, 
\end{eqnarray}
%
%
respectively. Under the transformation (4.17) and the finite gauge 
transformation ${\cal A}_{\mu\sigma} \rightarrow 
\bar{\cal A}_{\mu\sigma} 
= {\cal X}{\cal A}_{\mu\sigma}{\cal X}^{-1} 
-i{\cal X} \partial_{\mu\sigma} {\cal X}^{-1}$, we obtain the finite 
gauge transformation rule of $\widehat{\cal A}_{\mu\sigma}$: 
\begin{eqnarray}
\widehat{\cal A}_{\mu\sigma} \rightarrow 
\bar{\!\!\widehat{\cal A}}_{\mu\sigma} 
= {\cal Y}\widehat{\cal A}_{\mu\sigma}{\cal Y}^{-1} 
-i{\cal Y} \partial_{\mu\sigma} {\cal Y}^{-1} \;.
\end{eqnarray}
%
Consequently, the (naive) field strength 
$\widehat{\cal F}_{\mu\rho, \nu\sigma} 
\equiv \partial_{\mu\rho}\widehat{\cal A}_{\nu\sigma}
- \partial_{\nu\sigma}\widehat{\cal A}_{\mu\rho}
+ i[\,\widehat{\cal A}_{\mu\rho}, \widehat{\cal A}_{\nu\sigma}\,]$ obeys the 
transformation rule 
\begin{eqnarray}
\widehat{\cal F}_{\mu\rho, \nu\sigma} 
\rightarrow \bar{\!\widehat{\cal F}}_{\mu\rho, \nu\sigma}
= {\cal Y}\widehat{\cal F}_{\mu\rho, \nu\sigma}{\cal Y}^{-1} \;.
\end{eqnarray}
%
Replacing $\Xi^{(0)}$ with the infinitesimal function $\Lambda^{(0)}$ in 
${\cal Y}$, we obtain the infinitesimal transformation rule of 
$\widehat{\cal F}_{\mu\rho, \nu\sigma}$ from (4.22): 
\begin{eqnarray}
\delta \widehat{\cal F}_{\mu\rho, \nu\sigma} 
= i[\,\widehat{\cal F}_{\mu\rho, \nu\sigma}, \Lambda^{(0)} \,] \;.
\end{eqnarray}
%
As in (4.8), we express $\widehat{\cal F}_{\mu\rho, \nu\sigma}$ 
as a linear combination, $\widehat{\cal F}_{\mu\rho, \nu\sigma} = 
\sum^{\infty}_{p=0}\widehat{\cal F}^{(p)}_{\mu\rho, \nu\sigma}$, with
\begin{eqnarray}
\widehat{\cal F}^{(p)}_{\mu\rho, \nu\sigma} = 
\widehat{\cal F}^{Y(p)}_{\mu\rho, \nu\sigma} +
\widehat{\cal F}^{U(p)}_{\mu\rho, \nu\sigma} \;,
\end{eqnarray}
%
and
\begin{eqnarray}
& & \widehat{\cal F}^{Y(p)}_{\mu\rho, \nu\sigma} 
 = \partial_{\mu\rho}\widehat{\cal A}^{Y(p)}_{\nu\sigma} 
 - \partial_{\nu\sigma}\widehat{\cal A}^{Y(p)}_{\mu\rho}
 + i \sum^{p}_{k=0}[\,\widehat{\cal A}^{Y(k)}_{\mu\rho}, 
  \widehat{\cal A}^{Y(p-k)}_{\nu\sigma}\,]^{Y} \;, \\
 \nonumber \\
& & \widehat{\cal F}^{U(p)}_{\mu\rho, \nu\sigma} 
 = \partial_{\mu\rho}\widehat{\cal A}^{U(p)}_{\nu\sigma}
 - \partial_{\nu\sigma}\widehat{\cal A}^{U(p)}_{\mu\rho}
 + i \sum^{p}_{k=0}[\,\widehat{\cal A}^{Y(k)}_{\mu\rho} , 
      \widehat{\cal A}^{Y(p-k)}_{\nu\sigma}\,]^{U} \;.
\end{eqnarray}
%
Here, $\widehat{\cal F}^{(0)}_{\mu\rho, \nu\sigma}$ is identical with 
${\cal F}^{(0)}_{\mu\rho, \nu\sigma}$. From (4.23), we see that 
the infinitesimal transformation rules of 
$\widehat{\cal F}^{(p)}_{\mu\rho, \nu\sigma} \;\; (p=0,1,2,\ldots)$ 
are given by $\delta \widehat{\cal F}^{(p)}_{\mu\rho, \nu\sigma} 
= i[\,\widehat{\cal F}^{(p)}_{\mu\rho, \nu\sigma}, \Lambda^{(0)} \,]$. 
Note that every $\widehat{\cal F}^{(p)}_{\mu\rho, \nu\sigma} \;\; 
(p=0,1,2,\ldots)$ obeys the same transformation 
rule as $\widehat{\cal F}_{\mu\rho, \nu\sigma}$. From 
$\widehat{\cal F}^{(p)}_{\mu\rho, \nu\sigma}$, we can construct the modified 
field strengths of ${\cal A}^{(p)}_{\mu\sigma}\;(p=1,2,3,\ldots)$ 
in the manner discussed in Sec. 2.

We define the field strength of ${\cal A}^{(p)}_{\mu\sigma}$ 
in the same way as (2.14):
\begin{eqnarray} 
\widehat{\cal H}^{(p)}_{\mu\rho, \nu\sigma} 
= \widehat{\cal F}^{Y(p)}_{\mu\rho, \nu\sigma} 
+ \widehat{\cal H}^{U(p)}_{\mu\rho, \nu\sigma} \;,
\end{eqnarray}
%
with 
\begin{eqnarray}
\widehat{\cal H}^{U(p)}_{\mu\rho, \nu\sigma} 
\!\!&\equiv&\!\! \widehat{\cal F}^{U(p)}_{\mu\rho, \nu\sigma}
+ k\int\frac{d\omega}{2\pi}x'^{\lambda}(\omega)
{\rm Tr} \left[ {\cal A}^{Y(0)}_{\lambda\omega}
\widehat{\cal F}^{Y(p)}_{\mu\rho, \nu\sigma} \right] \;,
\end{eqnarray}
%
%
where $\widehat{\cal H}^{U(p)}_{\mu\rho, \nu\sigma}$ becomes 
gauge-invariant owing to the condition (2.7).

Finally, we obtain the action for ${\cal A}^{(p)}_{\mu\sigma}$ as 
\begin{eqnarray}
S_{\rm R}^{(p)} = \frac{1}{V_{R}} \int [dx] 
\left( {\cal L}^{Y(p)} + {\cal L}^{U(p)} \right) 
 \exp\left( -\frac{L}{l^{2}}\right) \;, 
\end{eqnarray}
%
with the Lagrangians
\begin{eqnarray}
& & {\cal L}^{Y(p)} = -\frac{1}{4}N^{Y}
 {\cal G}^{\kappa\rho, \lambda\sigma}{\cal G}^{\mu\chi, \nu\omega}
 {\rm Tr}[\,\widehat{\cal F}^{Y(p)}_{\kappa\rho, \mu\chi} \; 
 \widehat{\cal F}^{Y(p)}_{\lambda\sigma, \nu\omega} \,] \;, \\
 \nonumber \\
& & {\cal L}^{U(p)} = -\frac{1}{4} N^{U}
 {\cal G}^{\kappa\rho, \lambda\sigma}{\cal G}^{\mu\chi, \nu\omega}
 \widehat{\cal H}^{U(p)}_{\kappa\rho, \mu\chi}
 \widehat{\cal H}^{U(p)}_{\lambda\sigma, \nu\omega} \;,
\end{eqnarray}
%
where the same damping factor is introduced as in (2.17). Obviously, 
the Lagrangians ${\cal L}^{Y(p)}$ and ${\cal L}^{U(p)}$ are 
gauge invariant. Also, the action $S_{\rm R}^{(p=0)}$ is identical 
with the action given in Sec. 3. In this way, the suitable action for 
${\cal A}^{(p)}_{\mu\sigma}$ can be systematically derived using the 
formalism of the non-linear realization.


%
\section{The Chapline-Manton coupling for higher rank tensor fields}
\setcounter{equation}{0}

Referring to the previous section, 
we now derive the local field theory with couplings among 
non-abelian tensor fields of second rank, local Yang-Mills 
fields and abelian tensor fields of third rank. 
Let us consider the next simplest solutions (2.7) and (2.8) 
consisting of local functions on $M^{D}$, 
\begin{eqnarray}
\Lambda^{Y(1)}[x] \!\!&=&\!\! \int^{2\pi}_{0} \frac{d\sigma}{2\pi}
g_{1}Q^{\mu}(\sigma)\, \lambda_{\mu}{}^{a}(x(\sigma))\,T_{a}(\sigma) \;, \\
\nonumber \\
\Lambda^{U(1)}[x] \!\!&=&\!\! \int^{2\pi}_{0} \frac{d\sigma}{2\pi}
 q_{1}\sqrt{-x'^{\,2}(\sigma)} Q^{\mu}(\sigma)Q^{\nu}(\sigma)\,
 \xi_{\mu\nu}(x(\sigma)) 
\nonumber \\
\!\!&+&\!\! \int^{2\pi}_{0} \frac{d\sigma}{2\pi} 
 e_{1}Q'^{\mu}(\sigma)\,\kappa_{\mu}(x(\sigma)) \;, 
\end{eqnarray}
%
where $\lambda_{\mu}{}^{a}(x)$ and $\kappa_{\mu}(x)$ are 
infinitesimal vector functions, and $\xi_{\mu\nu}(x)$ is an 
infinitesimal tensor function of second rank. The constants 
$g_{1}$ and $e_{1}$ 
are of dimension $[\,{\rm length}\,]^{\frac{D-4}{2}}$, and $q_{1}$ is of 
dimension $[\,{\rm length}\,]^{\frac{D-6}{2}}$. The infinitesimal functions 
(5.1) and (5.2) correspond to (4.1) and (4.2) with $p=1$, respectively. 
In this case, however, we can rewrite (5.2) in a more simple form as 
\begin{eqnarray}
\Lambda^{U(1)}[x] = \int^{2\pi}_{0} \frac{d\sigma}{2\pi}
 q_{1}\sqrt{-x'^{\,2}(\sigma)} Q^{\mu}(\sigma)Q^{\nu}(\sigma)\,
 \lambda_{\mu\nu}(x(\sigma)) 
\end{eqnarray}
%
with a redefinition of the infinitesimal tensor function: 
$\lambda_{\mu\nu}(x) \equiv \xi_{\mu\nu}(x) 
- (e_{1}/2q_{1})\partial_{(\mu}\kappa_{\nu)}(x)$. 
In deriving (5.3), we carried out a partial integration over $\sigma$ for 
the second term on the right-hand side of (5.2). 
Corresponding to (5.1) and (5.3), we take the Yang-Mills field 
${\cal A}^{Y(1)}_{\mu\sigma}$ and the {\it U}(1) gauge field 
${\cal A}^{U(1)}_{\mu\sigma}$ consisting of local fields on $M^{D}$ as
\begin{eqnarray}
{\cal A}^{Y(1)}_{\mu\sigma}[x]
\!\!&=&\!\! g_{1} Q^{\nu}(\sigma)B_{\mu\nu}{}^{a}(x(\sigma))T_{a}(\sigma) 
 - \frac{1}{2} g_{1}Q_{\mu}(\sigma)Q^{\nu}(\sigma)Q^{\kappa}(\sigma)\,
 C_{\nu\kappa}{}^{a}(x(\sigma))T_{a}(\sigma) \nonumber \\
\!\!&-&\!\! \tilde{g}_{1} \phi_{\nu}{}^{a}(x(\sigma)) D_{\sigma}
 \left( \Pi_{\mu}{}^{\nu}(\sigma) \frac{T_{a}(\sigma)}{\sqrt{-x'^{2}(\sigma)}} 
 \right) \;, \\
\nonumber \\ 
\widetilde{{\cal A}}^{U(1)}_{\mu\sigma}[x] 
\!\!&=&\!\! q_{1} \sqrt{-x'^{2}(\sigma)}Q^{\nu}(\sigma)Q^{\kappa}(\sigma)
 U_{\mu\nu\kappa}(x(\sigma)) \nonumber \\ 
\!\!&+&\!\! q_{1} \sqrt{-x'^{2}(\sigma)}
 Q_{\mu}(\sigma)Q^{\nu}(\sigma)Q^{\kappa}(\sigma)Q^{\lambda}(\sigma) 
 V_{\nu\kappa\lambda}(x(\sigma)) \nonumber \\ 
\!\!&-&\!\! \tilde{q}_{1}Q'^{\nu}(\sigma)\{ B_{\mu\nu}(x(\sigma)) 
  + C_{\mu\nu}(x(\sigma)) \} \nonumber \\
\!\!&-&\!\! \tilde{q}_{1} \frac{d}{d\sigma}\left( 
 Q_{\mu}(\sigma)Q^{\nu}(\sigma)Q^{\kappa}(\sigma) \right)
 \phi_{\nu\kappa}(x(\sigma)) \;,
\end{eqnarray}
%
where $\Pi^{\mu}{}_{\nu}(\sigma) \equiv \delta^{\mu}{}_{\nu} + 
Q^{\mu}(\sigma)Q_{\nu}(\sigma)$. The differential $D_{\sigma}$ is defined as 
$D_{\sigma}P_{a}(\sigma) \equiv dP_{a}(\sigma)/d\sigma 
+ g_{0}x'^{\mu}(\sigma)A_{\mu}{}^{b}(x(\sigma))f_{ba}{}^{c}P_{c}(\sigma)$ 
with $P_{a}(\sigma)$ an arbitrary function of $\sigma$, where the 
$A_{\mu}{}^{a}(x)$ are the local Yang-Mills fields. 
Here, $U_{\mu\nu\lambda}(x)$ is a tensor field of third rank with 
the symmetric property $U_{\lambda\mu\nu}=U_{\lambda\nu\mu}$, 
$V_{\mu\nu\lambda}(x)$ is a totally symmetric tensor field of 
third rank, $B_{\mu\nu}{}^{a}(x)$ and $B_{\mu\nu}(x)$ are antisymmetric 
tensor fields of second rank, $C_{\mu\nu}{}^{a}(x)$, $C_{\mu\nu}(x)$ and 
$\phi_{\mu\nu}(x)$ are symmetric tensor fields of second rank, and 
$\phi_{\mu}{}^{a}(x)$ are vector fields. 
The constants 
$\tilde{g}_{1}$ and $\tilde{q}_{1}$ are of dimension 
$[{\rm length}]^{\frac{D-6}{2}}$ and $[{\rm length}]^{\frac{D-2}{2}}$, 
respectively.

From (4.7), the gauge-transformation rules of ${\cal A}^{Y(1)}_{\mu\sigma}$ 
and ${\cal A}^{U(1)}_{\mu\sigma}$ are given by
\begin{eqnarray}
& & \delta {\cal A}^{Y(1)}_{\mu\sigma} 
 = \partial_{\mu\sigma} \Lambda^{Y(1)} 
 + i [\, {\cal A}^{Y(1)}_{\mu\sigma}, \Lambda^{Y(0)} \,]^{Y} 
 + i [\, {\cal A}^{Y(0)}_{\mu\sigma}, \Lambda^{Y(1)} \,]^{Y} \;, \\
\nonumber \\
& & \delta \widetilde{{\cal A}}^{U(1)}_{\mu\sigma} 
 = \partial_{\mu\sigma} \Lambda^{U(1)} 
 + i [\, {\cal A}^{Y(1)}_{\mu\sigma}, \Lambda^{Y(0)} \,]^{U} 
 + i [\, {\cal A}^{Y(0)}_{\mu\sigma}, \Lambda^{Y(1)} \,]^{U} \;. 
\end{eqnarray}
%
%
Substituting (3.1), (3.3), (5.1) and (5.4) into (5.6), we obtain 
the infinitesimal transformation rules of the local fields 
$B_{\mu\nu}{}^{a}(x)$, $C_{\mu\nu}{}^{a}(x)$ and $\phi_{\mu}{}^{a}(x)$ as
\begin{eqnarray}
& & \delta B_{\mu\nu}{}^{a}(x) = D_{[\mu}\lambda_{\nu]}{}^{a}(x) 
   - g_{0}B_{\mu\nu}{}^{b}(x)\lambda^{c}(x)f_{bc}{}^{a} \;, \\
\nonumber \\
& & \delta C_{\mu\nu}{}^{a}(x) = D_{(\mu}\lambda_{\nu)}{}^{a}(x) 
   - g_{0}C_{\mu\nu}{}^{b}(x)\lambda^{c}(x)f_{bc}{}^{a} \;, \\
\nonumber \\
& & \delta \phi_{\mu}{}^{a}(x) = m_{g1}\lambda_{\mu}{}^{a}(x) 
   - g_{0}\phi_{\mu}{}^{b}(x)\lambda^{c}(x)f_{bc}{}^{a} \;,
\end{eqnarray}
%
where $m_{g1} \equiv g_{1}/\tilde{g}_{1}$ is a constant of dimension 
$[{\rm length}]^{-1}$ and $D_{\mu}$ denotes the covariant derivative given 
by (3.5). We see that the local fields $B_{\mu\nu}{}^{a}(x)$, 
$C_{\mu\nu}{}^{a}(x)$ and $\phi_{\mu}{}^{a}(x)$ obey non-abelian gauge 
transformation rules \cite{LSYM}. Similarly, substituting 
(3.1), (3.2), (3.3), (5.1), (5.3), (5.4) and (5.5) into (5.7) yields the 
infinitesimal transformation rules of the local fields $U_{\mu\nu\lambda}(x)$, 
$V_{\mu\nu\lambda}(x)$, $B_{\mu\nu}(x)$, $C_{\mu\nu}(x)$ and 
$\phi_{\mu\nu}(x)$:
\begin{eqnarray}
\delta B_{\mu\nu}(x) 
\!\!&=&\!\! \frac{\tilde{k}_{1}}{2}\left( 
m_{q1}A_{[\mu}{}^{a}(x)\lambda_{\nu], a}(x) 
- \phi_{[\mu}{}^{a}(x)\partial_{\nu]} \lambda_{a}(x) \right) \;, \\
\nonumber \\
\delta C_{\mu\nu}(x) 
\!\!&=&\!\! 2m_{q1}\lambda_{\mu\nu}(x) + \frac{\tilde{k}_{1}}{2}\left( 
m_{q1}A_{(\mu}{}^{a}(x)\lambda_{\nu), a}(x) 
- \phi_{(\mu}{}^{a}(x)\partial_{\nu)} \lambda_{a}(x) \right)\;, \\
\nonumber \\
\delta U_{\mu\nu\kappa}(x) 
\!\!&=&\!\! 
\partial_{\mu}\lambda_{\nu\kappa}(x) - 
\partial_{(\nu}\lambda_{\kappa)\mu}(x) 
\nonumber \\
\!\!&-&\!\! \frac{\tilde{k}_{1}}{2} \left( 
A_{\mu}{}^{a}(x)\partial_{(\nu}\lambda_{\kappa), a}(x) 
+ B_{\mu(\nu}{}^{a}\partial_{\kappa)}\lambda_{a}(x)
- \frac{1}{m_{g1}}\phi_{\mu}{}^{a}(x)D_{(\nu}\partial_{\kappa)}\lambda_{a}(x)
\right) \;, \nonumber \\
 \\
\delta V_{\mu\nu\kappa}(x) 
\!\!&=&\!\! 
-\frac{1}{6}\partial_{(\mu}\lambda_{\nu\kappa)}(x) 
\nonumber \\
\!\!&+&\!\! \frac{\tilde{k}_{1}}{12} \left( 
C_{(\mu\nu}{}^{a}\partial_{\kappa)}\lambda_{a}(x)
+ \frac{2}{m_{g1}}\phi_{(\mu}{}^{a}(x)D_{\nu}\partial_{\kappa)}\lambda_{a}(x)
\right) \;,  \footnotemark {}^{)}  \\
\nonumber \\
\delta \phi_{\mu\nu}(x) \!\!&=&\!\! m_{q1}\lambda_{\mu\nu}(x)
  - \frac{\tilde{k}_{1}}{2}\frac{m_{q1}}{m_{g1}}
  \phi_{(\mu}{}^{a}(x)\partial_{\nu)}\lambda_{a}(x)  \;.
\end{eqnarray}
%
Here $m_{q1} \equiv q_{1}/\tilde{q}_{1}$ and $\tilde{k}_{1} \equiv 
kg_{0}g_{1}/q_{1}$ are constants of dimension $[{\rm length}]^{-1}$ 
and $[{\rm length}]$, respectively. 
The local fields $U_{\mu\nu\lambda}(x)$, $V_{\mu\nu\lambda}(x)$, 
$B_{\mu\nu}(x)$, $C_{\mu\nu}(x)$ and $\phi_{\mu\nu}(x)$ obey abelian 
gauge transformation rules. As we expected, the non-abelian local fields 
$A_{\mu}{}^{a}$, $B_{\mu\nu}{}^{a}$, $C_{\mu\nu}{}^{a}$ and $\phi_{\mu}{}^{a}$ 
appear in the infinitesimal transformation rules of these abelian local fields.

As in the case of (3.8), we can find the gauge invariant tensor fields 
under the transformation rules (5.8)--(5.15) and (3.5).  We obtain 
\footnotetext{
${}^{)}$ 
$X_{(\mu}Y_{\nu}Z_{\lambda)} \equiv 
X_{\mu}Y_{\nu}Z_{\lambda}+X_{\nu}Y_{\lambda}Z_{\nu}+X_{\lambda}Y_{\mu}Z_{\nu}
+X_{\lambda}Y_{\nu}Z_{\mu}+X_{\nu}Y_{\mu}Z_{\lambda}+X_{\mu}Y_{\lambda}Z_{\nu}$ . }
%
\begin{eqnarray}
& & \widetilde{B}_{\mu\nu}(x) \equiv 
  B_{\mu\nu}(x) 
  - \frac{\tilde{k}_{1}}{2}\frac{m_{q1}}{m_{g1}}
  A_{[\mu}{}^{a}(x)\phi_{\nu], a}(x)
  \;, \\
\nonumber \\
& & \widetilde{C}_{\mu\nu}(x) \equiv
 C_{\mu\nu}(x) - 2\phi_{\mu\nu}(x)
 - \frac{\tilde{k}_{1}}{2}\frac{m_{q1}}{m_{g1}}
   A_{(\mu}{}^{a}(x)\phi_{\nu), a}(x)
 \;, \\
\nonumber \\
& & \widetilde{V}_{\mu\nu\lambda}(x) \equiv
 V_{\mu\nu\lambda}(x) - S_{\mu\nu\lambda}(x)
 - \frac{\tilde{k}_{1}}{12} C_{(\mu\nu}{}^{a}(x)A_{\lambda), a}(x)
  \;, 
\end{eqnarray}
%
where $S_{\mu\nu\lambda}(x)$ is the irreducible component of 
$U_{\mu\nu\lambda}(x)$ given by $S_{\mu\nu\lambda} \equiv 
( U_{\mu\nu\lambda} + U_{\nu\lambda\mu}+U_{\lambda\mu\nu} )/3$.

Eliminating $B_{\mu\nu}$, $C_{\mu\nu}$ and 
$V_{\mu\nu\lambda}$ from (5.5) by using (5.16), (5.17) and 
(5.18), we can rewrite (5.5) in terms of the abelian local fields 
$\widetilde{B}_{\mu\nu}$, $\widetilde{C}_{\mu\nu}$, 
$\widetilde{V}_{\mu\nu\lambda}$, 
$U_{\mu\nu\lambda}$, $S_{\mu\nu\lambda}$ 
and $\phi_{\mu\nu}$ and the non-abelian local fields 
$C_{\mu\nu}{}^{a}$, $\phi_{\mu}{}^{a}$ and 
$A_{\mu}{}^{a}$. As in Sec. 3, 
we next remove the gauge invariant tensor fields 
$\widetilde{B}_{\mu\nu}$, $\widetilde{C}_{\mu\nu}$ and 
$\widetilde{V}_{\mu\nu\lambda}$ from (5.5). Then (5.5) 
can be rewritten as 
\begin{eqnarray}
\widetilde{{\cal A}}^{U(1)}_{\mu\sigma}[x]
\!\!&=&\!\! {\cal A}^{U(1)}_{\mu\sigma}[x]
 - \tilde{k}_{1}q_{1}\frac{1}{m_{g1}}
 Q'^{\nu}(\sigma)A_{\mu}{}^{a}(x(\sigma))\phi_{\nu, a}(x(\sigma)) 
\nonumber \\
\!\!&+&\!\! \frac{\tilde{k}_{1}}{2}q_{1}\sqrt{-x'^{2}(\sigma)}
 Q_{\mu}(\sigma)Q^{\nu}(\sigma)Q^{\kappa}(\sigma)Q^{\lambda}(\sigma)
 C_{\nu\kappa}{}^{a}(x(\sigma))A_{\lambda, a}(x(\sigma)) \;,
\nonumber \\
\end{eqnarray}
%
with
\begin{eqnarray}
{\cal A}^{U(1)}_{\mu\sigma}[x]
\!\!&=&\!\! q_{1} \sqrt{-x'^{2}(\sigma)}
 \{ \delta_{\mu}{}^{[\nu} Q^{\kappa]}(\sigma)Q^{\zeta}(\sigma) 
  - \Pi_{\mu}{}^{\zeta}(\sigma)Q^{\nu}(\sigma)Q^{\kappa}(\sigma) \}
 A_{\nu\kappa\zeta}(x(\sigma)) \nonumber \\
\!\!&-&\!\! \tilde{q}_{1} \sqrt{-x'^{2}(\sigma)}
 \{ Q'_{\mu}(\sigma)Q^{\nu}(\sigma)Q^{\kappa}(\sigma)
  + 2\Pi_{\mu}{}^{\nu}(\sigma)Q'^{\kappa}(\sigma) \}
 \phi_{\nu\kappa}(x(\sigma)) \;,
\nonumber \\
\end{eqnarray}
%
where $A_{\mu\nu\lambda}(x)$ is a tensor field of third rank defined by 
$A_{\mu\nu\lambda} \equiv (U_{\mu\nu\lambda}-3S_{\mu\nu\lambda})/2$ and has 
the symmetry property $A_{\lambda\mu\nu}=A_{\lambda\nu\mu}$. The 
infinitesimal transformation rule of $A_{\mu\nu\lambda}(x)$ is given by
\begin{eqnarray}
\delta A_{\mu\nu\kappa}(x) 
\!\!&=&\!\! 
 \partial_{\mu}\lambda_{\nu\kappa}(x) \nonumber \\
\!\!&+&\!\!
  \frac{\tilde{k}_{1}}{4}  
  \Biggl( A_{(\nu}{}^{a}(x)\partial_{\kappa)}\lambda_{\mu, a}(x)
  + \partial_{\mu}\lambda_{(\nu}{}^{a}(x)A_{\kappa), a}(x) 
  - B_{\mu(\nu}{}^{a}(x)\partial_{\kappa)}\lambda_{a}(x)
  \Biggr. \nonumber \\
& & - \Biggl. 
  \frac{1}{m_{g1}}\phi_{(\nu}{}^{a}(x) 
  D_{\kappa)}\partial_{\mu}\lambda_{a}(x)
  - \frac{1}{m_{g1}}D_{\mu}\partial_{(\nu}\lambda^{a}(x)\phi_{\kappa), a}(x)
  \Biggr) \;.
\end{eqnarray}
%
The {\it U}(1) gauge field $\widetilde{\cal A}^{U(1)}_{\mu\sigma}$ is 
reduced to ${\cal A}^{U(1)}_{\mu\sigma}$ by setting $k=0$. 
As we have seen in Sec. 3, we can also check that 
$\widetilde{\cal A}^{U(1)}_{\mu\sigma}$ does not 
satisfy the condition 
$x'^{\mu}(\sigma)\widetilde{\cal A}^{U(1)}_{\mu\sigma}=0$ if $k \neq 0$, 
while ${\cal A}^{U(1)}_{\mu\sigma}$, which is written in terms 
of the abelian local tensor fields, satisfies the 
condition $x'^{\mu}(\sigma){\cal A}^{U(1)}_{\mu\sigma}=0$. \cite{LSU(1)T}

Next, let us consider the scalar field 
$\Phi^{(1)} = \Phi^{Y(1)} + \Phi^{U(1)}$, 
where $\Phi^{(1)}$ corresponds to (4.16) with $p=1$. From (4.17), we see that 
$\Phi^{Y(1)}$ and $\Phi^{U(1)}$ obey the infinitesimal transformation 
rules 
\begin{eqnarray}
& & \delta \Phi^{Y(1)} 
= \Lambda^{Y(1)} + i[\,\Phi^{Y(1)}, \Lambda^{Y(0)} \,]^{Y} \;, \\
\nonumber \\ 
& & \delta \Phi^{U(1)} 
= \Lambda^{U(1)} + i[\,\Phi^{Y(1)}, \Lambda^{Y(0)} \,]^{U} \;.
\end{eqnarray}
We express the scalar fields $\Phi^{Y(1)}$ and $\Phi^{U(1)}$ in terms of the 
local fields on $M^{D}$. Since the infinitesimal scalar functions 
$\Lambda^{Y(1)}$, $\Lambda^{U(1)}$ and $\Lambda^{Y(0)}$ are 
reparametrization invariant, $\Phi^{Y(1)}$ and $\Phi^{U(1)}$ also have to be 
reparametrization invariant. Taking account of this, we express $\Phi^{Y(1)}$ 
and $\Phi^{U(1)}$ in terms of the local fields 
$\phi_{\mu}{}^{a}$ and $\phi_{\mu\nu}$, respectively, as 
\begin{eqnarray}
& & \Phi^{Y(1)}[x] = \int^{2\pi}_{0} \frac{d\sigma}{2\pi} \tilde{g}_{1} 
 Q^{\mu}(\sigma) \phi_{\mu}{}^{a}(x(\sigma)) T_{a}(\sigma) \;, \\
\nonumber \\
& & \Phi^{U(1)}[x] = \int \frac{d\sigma}{2\pi} \tilde{q}_{1}
 \sqrt{-x'^{2}(\sigma)} Q^{\mu}(\sigma)Q^{\nu}(\sigma) 
 \phi_{\mu\nu}(x(\sigma)) \;,
\end{eqnarray}
%
where $\tilde{g}_{1}$ and $\tilde{q}_{1}$ are the constants appearing in (5.4) 
and (5.5). The transformation rules (5.22) and (5.23) lead to infinitesimal 
transformation rules of $\phi_{\mu\nu}$ and 
$\phi_{\mu}{}^{a}$. However, we find that the transformation rules of 
$\phi_{\mu}{}^{a}$ and $\phi_{\mu\nu}$ are compatible with (5.10) 
and (5.15).

From (4.20),  the vector fields $\widehat{\cal A}^{Y(1)}_{\mu\sigma}$ and 
$\widehat{\cal A}^{U(1)}_{\mu\sigma}$ are given by 
\begin{eqnarray}
& & \widehat{{\cal A}}^{Y(1)}_{\mu\sigma}
 \equiv {\cal A}^{Y(1)}_{\mu\sigma} - \partial_{\mu\sigma}\Phi^{Y(1)} 
 + i[\,{\cal A}^{Y(0)}_{\mu\sigma}, \Phi^{Y(1)} \,]^{Y} \;, \\
\nonumber \\
& & \widehat{{\cal A}}^{U(1)}_{\mu\sigma}
 \equiv \widetilde{\cal A}^{U(1)}_{\mu\sigma} - \partial_{\mu\sigma}\Phi^{U(1)}  
+ i[\,{\cal A}^{Y(0)}_{\mu\sigma}, \Phi^{Y(1)} \,]^{U} \;. 
\end{eqnarray}
%
Substituting (5.4), (5.24), and (3.3) into (5.26), we obtain 
\begin{eqnarray}
\widehat{{\cal A}}^{Y(1)}_{\mu\sigma} 
\!\!&=&\!\! 
g_{1} Q^{\nu}(\sigma) \widehat{B}_{\mu\nu}{}^{a}(x(\sigma))T_{a}(\sigma) 
\nonumber \\
\!\!&-&\!\! 
 \frac{1}{2}g_{1}Q_{\mu}(\sigma)Q^{\nu}(\sigma)Q^{\lambda}(\sigma)
 \widehat{C}_{\nu\lambda}{}^{a}(x(\sigma))T_{a}(\sigma) \;, 
\end{eqnarray}
%
with
\begin{eqnarray}
\widehat{B}_{\mu\nu}{}^{a}(x) \!\!&\equiv&\!\! B_{\mu\nu}{}^{a}(x)
   - \frac{1}{m_{g1}}D_{[\mu}\phi_{\nu]}{}^{a}(x) \;, \\
\widehat{C}_{\mu\nu}{}^{a}(x) \!\!&\equiv&\!\! C_{\mu\nu}{}^{a}(x)
   - \frac{1}{m_{g1}}D_{(\mu}\phi_{\nu)}{}^{a}(x) \;. 
\end{eqnarray}
%
Under the transformation rules (5.8), (5.9), (5,10) and (3.5), 
$\widehat{B}_{\mu\nu}{}^{a}$ and $\widehat{C}_{\mu\nu}{}^{a}$ transform 
homogeneously: 
\begin{eqnarray}
& & \delta \widehat{B}_{\mu\nu}{}^{a}(x) 
= -g_{0}\widehat{B}_{\mu\nu}{}^{b}(x)\lambda^{c}(x)f_{bc}{}^{a} \;, \\
 \nonumber \\
& & \delta \widehat{C}_{\mu\nu}{}^{a}(x) 
= -g_{0}\widehat{C}_{\mu\nu}{}^{b}(x)\lambda^{c}(x)f_{bc}{}^{a} \;. 
\end{eqnarray}
%
The transformation rules (5.31) and (5.32) are compatible with the 
gauge transformation rules $\delta \widehat{{\cal A}}^{Y(1)}_{\mu\sigma} 
= i[\,\widehat{{\cal A}}^{Y(1)}_{\mu\sigma}, \Lambda^{Y(0)} \,]^{Y}$. Also, 
substitution of (5,19), (5,24), (5.25) and (3.3) into  (5.27) yields 
\begin{eqnarray}
\widehat{{\cal A}}^{U(1)}_{\mu\sigma}[x] 
\!\!&=&\!\! q_{1}\sqrt{-x'^{2}(\sigma)} 
 Q^{\nu}(\sigma)Q^{\lambda}(\sigma)B_{\mu\nu\lambda}(x(\sigma)) \nonumber \\
\!\!&-&\!\! \frac{1}{2}q_{1}\sqrt{-x'^{2}(\sigma)}
 Q_{\mu}(\sigma)Q^{\nu}(\sigma)Q^{\kappa}(\sigma)Q^{\lambda}(\sigma)
 C_{\nu\kappa\lambda}(x(\sigma)) \;.
\end{eqnarray}
%
with
\begin{eqnarray}
B_{\mu\nu\lambda}(x) 
\!\!&\equiv&\!\! \widetilde{A}_{\mu\nu\lambda}(x) 
 - \widetilde{A}_{(\nu\lambda)\mu}(x)
 + \frac{\tilde{k}_{1}}{2m_{g1}}
  A_{\mu}{}^{a}(x)\partial_{(\nu}\phi_{\lambda), a}(x) \;, \\
C_{\mu\nu\lambda}(x) 
\!\!&\equiv&\!\! \frac{1}{3} \widetilde{A}_{(\mu\nu\lambda)}(x)
 - \frac{\tilde{k}_{1}}{6} C_{(\mu\nu}{}^{a}(x)A_{\lambda), a}(x) \;,
\end{eqnarray}
%
and $\widetilde{A}_{\mu\nu\lambda}(x) \equiv A_{\mu\nu\lambda}(x) 
- (1/m_{q1}) \partial_{\mu}\phi_{\nu\lambda}(x)$. Here, 
$\widetilde{A}_{\mu\nu\lambda}$ is a third-rank tensor field 
that is gauge invariant under $k=0$. \cite{LSU(1)T} 
By setting $k=0$, the abelian tensor 
fields $B_{\mu\nu\lambda}$ and $C_{\mu\nu\lambda}$ are reduced to the 
(irreducible) components of $\widetilde{A}_{\mu\nu\lambda}$. Under the 
transformation rules of (5.9), (5.10), (5.15), (5,21) and (3.5), we see that 
$B_{\mu\nu\lambda}$ and $C_{\mu\nu\lambda}$ obey simple transformation rules: 
\begin{eqnarray}
& & \delta B_{\mu\nu\kappa}(x) = -\frac{\tilde{k}_{1}}{2}
 \widehat{B}_{\mu(\nu}{}^{a}(x)\partial_{\kappa)}\lambda_{a}(x) \;, \\
\nonumber \\
& & \delta C_{\mu\nu\kappa}(x) = -\frac{\tilde{k}_{1}}{6}
 \widehat{C}_{(\mu\nu}{}^{a}(x)\partial_{\kappa)}\lambda_{a}(x) \;. 
\end{eqnarray}
%
These transformation rules are also compatible with the gauge transformation 
rule $\delta \widehat{{\cal A}}^{U(1)}_{\mu\sigma} 
 = i[\,\widehat{{\cal A}}^{Y(1)}_{\mu\sigma}, \Lambda^{Y(0)} \,]^{U}$.

We next express the field strength 
$\widehat{{\cal H}}^{(1)}_{\mu\rho, \nu\sigma}$ in terms of the local fields, 
where $\widehat{{\cal H}}^{(1)}_{\mu\rho, \nu\sigma}$ corresponds to (4.27) 
with $p=1$. The concrete forms of 
$\widehat{\cal F}^{Y(1)}_{\mu\rho, \nu\sigma}$ and 
$\widehat{\cal H}^{U(1)}_{\mu\rho, \nu\sigma}$ are given by 
\begin{eqnarray}
\widehat{{\cal F}}^{Y(1)}_{\mu\rho, \nu\sigma} 
\!\!&\equiv&\!\!
 \partial_{\mu\rho} \widehat{\cal A}^{Y(1)}_{\nu\sigma}
-\partial_{\nu\sigma} \widehat{\cal A}^{Y(1)}_{\mu\rho}
+ i[\,{\cal A}^{Y(0)}_{\mu\rho}, \widehat{\cal A}^{Y(1)}_{\nu\sigma}\,]^{Y} 
+ i[\,\widehat{\cal A}^{Y(1)}_{\mu\rho}, {\cal A}^{Y(0)}_{\nu\sigma}\,]^{Y} 
\;,   \nonumber \\
 \\
\widehat{{\cal H}}^{U(1)}_{\mu\rho, \nu\sigma} \!\!&\equiv&\!\!
 \partial_{\mu\rho} \widehat{{\cal A}}^{U(1)}_{\nu\sigma}
-\partial_{\nu\sigma} \widehat{{\cal A}}^{U(1)}_{\mu\rho}
+ i[\,{\cal A}^{Y(0)}_{\mu\rho}, \widehat{{\cal A}}^{Y(1)}_{\nu\sigma}\,]^{U} 
+ i[\,\widehat{{\cal A}}^{Y(1)}_{\mu\rho}, {\cal A}^{Y(0)}_{\nu\sigma}\,]^{U} 
\nonumber \\
\!\!&+&\!\! k \int\frac{d\omega}{2\pi}x'^{\lambda}(\omega){\rm Tr}
 \left[ {\cal A}^{Y(0)}_{\lambda\omega} 
 \widehat{{\cal F}}^{Y(1)}_{\mu\rho, \nu\sigma} \right] \;.
\end{eqnarray}
%
As was mentioned in Sec. 4, $\widehat{\cal F}^{Y(1)}_{\mu\rho, \nu\sigma}$ 
obeys the homogeneous gauge transformation rule 
$\delta \widehat{\cal F}^{Y(1)}_{\mu\rho, \nu\sigma} = 
i[\,\widehat{\cal F}^{Y(1)}_{\mu\rho, \nu\sigma}, \Lambda^{Y(0)}\,]^{Y}$, 
while $\widehat{\cal H}^{U(1)}_{\mu\rho, \nu\sigma}$ is gauge invariant: 
$\delta \widehat{\cal H}^{U(1)}_{\mu\rho, \nu\sigma} = 0$. 
Substituting (5.28) and (3.3) into (5.38), we obtain 
$\widehat{{\cal F}}^{Y(1)}_{\mu\rho, \nu\sigma}$ in terms of the local 
fields: 
\begin{eqnarray}
\widehat{{\cal F}}^{Y(1)}_{\mu\rho,\nu\sigma}[x] 
\!\!&=&\!\! \frac{1}{2}g_{1} \delta(\rho-\sigma) \Biggl{[} \Biggr.
 Q^{\lambda}(\rho)\widehat{H}_{\lambda\mu\nu}{}^{a}(x(\rho))T_{a}(\rho)
+ \widehat{B}_{\mu\nu}{}^{a}(x(\rho))D_{\rho} 
 \left( \frac{T_{a}(\rho)}{\sqrt{-x'^{2}(\rho)}} \right)
\nonumber \\
& &-\frac{1}{2}\widehat{I}_{\lambda[\mu}{}^{a}(x(\rho))D_{\rho} 
 \left( Q_{\nu]}(\rho)Q^{\lambda}(\rho)
 \frac{T_{a}(\rho)}{\sqrt{-x'^{2}(\rho)}} \right)
\nonumber \\
& &-\frac{1}{2} Q^{\kappa}(\rho)Q^{\lambda}(\rho)Q_{[\mu}(\rho) 
\widehat{J}_{\nu]\kappa\lambda}{}^{a}(x(\rho))T_{a}(\rho) \Biggl. \Biggr{]}
\nonumber \\
& &-\frac{1}{2}g_{1}\delta'(\rho-\sigma) \Biggl{[} \Biggr.
\{ Q^{\lambda}(\rho) Q_{(\mu}(\rho) 
\widehat{I}_{\nu)\lambda}{}^{a}(x(\rho))  \nonumber \\
& &-(3Q_{\mu}(\rho)Q_{\nu}(\rho)
 + \eta_{\mu\nu})Q^{\kappa}(\rho)Q^{\lambda}(\rho)
\widehat{C}_{\kappa\lambda}{}^{a}(x(\rho)) \} 
 \frac{T_{a}(\rho)}{\sqrt{-x'^{2}(\rho)}}
\Biggl. \Biggr{]} 
\nonumber \\
\nonumber \\
& & - (\mbox{\rm{all of the above terms with $\mu \leftrightarrow \nu$ and 
$\rho \leftrightarrow \sigma$. }}) \;, 
\end{eqnarray}
%
%
with
\begin{eqnarray}
\widehat{H}_{\lambda\mu\nu}{}^{a}(x) 
\!\!&\equiv&\!\!
 D_{\lambda}\widehat{B}_{\mu\nu}{}^{a}(x) 
+ D_{\mu}\widehat{B}_{\nu\lambda}{}^{a}(x)
+ D_{\nu}\widehat{B}_{\lambda\mu}{}^{a}(x)
\;, \\
 \nonumber \\
\widehat{I}_{\mu\nu}{}^{a}(x)
\!\!&\equiv&\!\!
\widehat{B}_{\mu\nu}{}^{a}(x) 
- \widehat{C}_{\mu\nu}{}^{a}(x) \;, \\
 \nonumber \\
\widehat{J}_{\lambda\mu\nu}{}^{a}(x)
\!\!&\equiv&\!\!
D_{\mu}\widehat{B}_{\lambda\nu}{}^{a}(x)
+ D_{\mu}\widehat{C}_{\lambda\nu}{}^{a}(x)
+ D_{\lambda}\widehat{C}_{\mu\nu}{}^{a}(x) \;.
\end{eqnarray}
%
It is obvious that $\widehat{H}_{\lambda\mu\nu}{}^{a}(x)$, 
$\widehat{I}_{\mu\nu}{}^{a}(x)$ and $\widehat{J}_{\lambda\mu\nu}{}^{a}(x)$ 
transform homogeneously. We next substitute (5.28), (5.33), (5.40) and 
(3.3) into (5.39). After a little tedious calculation, we obtain 
$\widehat{{\cal H}}^{U(1)}_{\mu\rho, \nu\sigma}$ in terms of 
the local fields: 
\begin{eqnarray}
& & \widehat{{\cal H}}^{U(1)}_{\mu\rho, \nu\sigma}
 = - \frac{1}{4}q_{1} \delta(\rho - \sigma) 
\nonumber \\
& & \times \Biggl{[} \Biggr.
 \sqrt{-x'^{2}(\rho)} \Biggl\{ \Biggr.
 - 2Q^{\kappa}(\rho)Q^{\lambda}(\rho)\delta_{[\dot{\mu}}{}^{\zeta} 
 + 2Q^{\zeta}(\rho)Q^{\lambda}(\rho)\delta_{[\dot{\mu}}{}^{\kappa} 
\nonumber \\
& & \qquad \qquad 
 + Q^{\kappa}(\rho)Q^{\lambda}(\rho)Q^{\zeta}(\rho)Q_{[\dot{\mu}}(\rho)
 \Biggl. \Biggr\} 
 \partial_{\zeta}\widetilde{B}_{\dot{\nu}]\kappa\lambda}(x(\rho)) 
\nonumber \\
& & + \sqrt{-x'^{2}(\sigma)} Q^{\lambda}(\rho)Q^{\zeta}(\rho) 
 \Biggl\{ \Biggr.
 - Q^{\kappa}(\rho)Q_{[\mu}(\rho)\delta_{\nu]}{}^{\omega} 
\nonumber \\
& & \qquad \qquad 
 + \frac{3}{2}Q^{\omega}(\rho)Q_{[\mu}(\rho)\delta_{\nu]}{}^{\kappa}
 \Biggl. \Biggr\} 
 \partial_{\omega}\widetilde{C}_{\kappa\lambda\zeta}(x(\rho)) 
\nonumber \\
& & + 2\tilde{k}_{1} \sqrt{-x'^{2}(\rho)}
 Q^{\kappa}(\rho)Q^{\lambda}(\rho)
 \widehat{B}_{\kappa[\mu}{}^{a}(x(\rho))F_{\nu]\lambda, a}(x(\rho)) 
\nonumber \\
& & + \tilde{k}_{1} \sqrt{-x'^{2}(\rho)}
 Q^{\kappa}(\rho)Q^{\lambda}(\rho)Q^{\zeta}(\rho)Q_{[\dot{\mu}}(\rho)
 \widehat{C}_{\kappa\lambda}{}^{a}(x(\rho)) F_{\dot{\nu}]\zeta, a}(x(\rho)) 
\nonumber \\
& & +\frac{d}{d\rho}
 \Biggl\{ \Biggr.
 2Q^{\lambda}(\rho)\delta_{[\mu}{}^{\kappa}
 + Q^{\kappa}(\rho)Q^{\lambda}(\rho)Q_{[\mu}(\rho)
 \Biggl. \Biggr\}
 \widetilde{B}_{\nu]\kappa\lambda}(x(\rho)) 
\nonumber \\
& & + \frac{3}{2}\frac{d}{d\rho}
 \Biggl\{ \Biggr.
 Q^{\kappa}(\rho)Q^{\lambda}(\rho)Q_{[\mu}(\rho)\delta_{\nu]}{}^{\zeta}
 \Biggl. \Biggr\}
 \widetilde{C}_{\kappa\lambda\zeta}(x(\rho))
 \Biggl. \Biggr{]}  
\nonumber \\
& & + \frac{1}{2}q_{1} \delta'(\rho - \sigma)  
\nonumber \\
& & \times \Biggl{[} \Biggr.
 - \Biggl\{ \Biggr.
 2Q^{\lambda}(\rho)\delta_{(\mu}{}^{\kappa}
 + Q^{\kappa}(\rho)Q^{\lambda}(\rho)Q_{(\mu}(\rho)
 \Biggl. \Biggr\}
 \widetilde{B}_{\nu)\kappa\lambda}(x(\rho)) 
\nonumber \\
& & + \Biggl\{ \Biggr.
 \left( 3Q_{\mu}(\rho)Q_{\nu}(\rho) + \eta_{\mu\nu} \right)
 Q^{\kappa}(\rho)Q^{\lambda}(\rho)Q^{\zeta}(\rho) 
\nonumber \\
& & \qquad \qquad + \frac{3}{2}
 Q^{\kappa}(\rho)Q^{\lambda}(\rho)Q_{(\mu}(\rho)\delta_{\nu)}{}^{\zeta}
 \Biggl. \Biggr\}
 \widetilde{C}_{\kappa\lambda\zeta}(x(\rho))
 \Biggl. \Biggr{]} \;\;
\footnotemark {}^{)}
\nonumber \\
\nonumber \\
& & - (\mbox{\rm{all of the above terms with $\mu \leftrightarrow \nu$ and 
$\rho \leftrightarrow \sigma$. }}) \;,
\end{eqnarray}
%
%
\footnotetext{${}^{)}$ 
$X_{[\dot{\mu}}Y_{\nu}Z_{\dot{\kappa}]} \equiv 
X_{\mu}Y_{\nu}Z_{\kappa}-X_{\kappa}Y_{\nu}Z_{\mu}, \qquad \quad 
X_{(\dot{\mu}}Y_{\nu}Z_{\dot{\kappa})} \equiv 
X_{\mu}Y_{\nu}Z_{\kappa}+X_{\kappa}Y_{\nu}Z_{\mu}$. }
%
%
with
\begin{eqnarray}
\widetilde{B}_{\mu\nu\lambda}(x)
\!\!&\equiv&\!\! 
  B_{\mu\nu\lambda}(x) + \frac{1}{2} \tilde{k}_{1} 
  \widehat{B}_{\mu(\nu}{}^{a}(x)A_{\lambda), a}(x) \;, \\
 \nonumber \\
\widetilde{C}_{\mu\nu\lambda}(x) 
\!\!&\equiv&\!\! 
 C_{\mu\nu\lambda}(x) + \frac{1}{6}\tilde{k}_{1} 
  \widehat{C}_{(\mu\nu}{}^{a}(x)A_{\lambda), a}(x) \;.
\end{eqnarray}
%
Here $F_{\mu\nu}{}^{a}$ is the field strength of $A_{\mu}{}^{a}$. 
Obviously, $\widetilde{B}_{\mu\nu\lambda}$ and 
$\widetilde{C}_{\mu\nu\lambda}$ are invariant under the transformation rules 
(5.31), (5.32), (5.36), (5.37) and (3.5). 
Besides $\widetilde{B}_{\mu\nu\lambda}$ and $\widetilde{C}_{\mu\nu\lambda}$ 
and their derivatives, the two distinctive terms  
$\widehat{B}_{\mu\nu}{}^{a}F_{\lambda\kappa, a}$ and 
$\widehat{C}_{\mu\nu}{}^{a}F_{\lambda\kappa, a}$ occur in (5.44). 
These terms are also invariant under the transformation rules of (5.31), 
(5.32) and (3.5). Consequently, 
$\widehat{\cal H}^{U(1)}_{\mu\rho, \nu\sigma}$ remains 
invariant under the transformation of the local fields. 
This invariance is compatible with the fact that 
$\delta \widehat{\cal H}^{U(1)}_{\mu\rho, \nu\sigma} = 0$. 
Note that the terms $\widehat{B}_{\mu\nu}{}^{a}F_{\lambda\kappa, a}$ 
and $\widehat{C}_{\mu\nu}{}^{a}F_{\lambda\kappa, a}$ 
take the form of products of the field 
strengths $F_{\mu\nu}{}^{a}$ and the non-abelian tensor fields 
$\widehat{B}_{\mu\nu}{}^{a}$ or $\widehat{C}_{\mu\nu}{}^{a}$.
These terms take the same form as the (non-abelian) 
{\it BF}-term, except for a totally antisymmetric property. \cite{BF} 
Accordingly, we refer to these terms as 
``{\it BF}-like terms'' hereafter. 
We can regard the {\it BF}-like terms in (5.44) as a kind of generalization 
of the Chern-Simons terms $\Omega_{\mu\nu\lambda}$ in (3.13) for the 
non-abelian tensor fields 
$\widehat{B}_{\mu\nu}{}^{a}$ and $\widehat{C}_{\mu\nu}{}^{a}$.

Finally, we consider the action $S_{{\rm R}}^{(1)}$ corresponding to (4.29) 
with $p=1$. We divide $S_{{\rm R}}^{(1)}$ into  $S^{Y(1)}_{{\rm R}}$ and 
$S^{U(1)}_{{\rm R}}$ as 
\begin{eqnarray}
S^{Y(1)}_{{\rm R}} 
\!\!&=&\!\! \frac{1}{V_{R}} \int [dx] {\cal L}^{Y(1)} 
\exp \left( -\frac{L}{l^{2}} \right) \;, \\
\nonumber \\
S^{U(1)}_{{\rm R}} 
\!\!&=&\!\! \frac{1}{V_{R}} \int [dx] {\cal L}^{U(1)} 
\exp \left( -\frac{L}{l^{2}} \right) \;, 
\end{eqnarray}
%
with the Lagrangians
\begin{eqnarray}
& & {\cal L}^{Y(1)} = - \frac{1}{4} N^{Y} 
{\cal G}^{\kappa\rho, \lambda\sigma}{\cal G}^{\mu\chi, \nu\omega}
{\rm Tr} [ \widehat{{\cal F}}^{Y(1)}_{\kappa\rho, \mu\chi}
\widehat{{\cal F}}^{Y(1)}_{\lambda\sigma, \nu\omega} ] \;, \\
 \nonumber \\
& & {\cal L}^{U(1)} = - \frac{1}{4} N^{U}
{\cal G}^{\kappa\rho, \lambda\sigma}{\cal G}^{\mu\chi, \nu\omega}
\widehat{{\cal H}}^{U(1)}_{\kappa\rho, \mu\chi}
\widehat{{\cal H}}^{U(1)}_{\lambda\sigma, \nu\omega} \;. 
\end{eqnarray}
%
From the definition of (5.47), it is obvious that 
$S^{Y(1)}_{{\rm R}}$ does not include the central charge $k$. This means that 
the action $S^{Y(1)}_{{\rm R}}$ is not affected by the central 
extension of the gauge group. 
In other words, the action $S^{Y(1)}_{{\rm R}}$ 
is coincident with the action whose gauge group is the 
loop group. 
The action $S^{Y(1)}_{{\rm R}}$ written in terms of 
the local fields is derived in Ref. 14). 
It describes a massive tensor field theory as 
non-abelian St\"uckelberg formalism for the tensor fields 
$\widehat{B}_{\mu\nu}{}^{a}$ and $\widehat{C}_{\mu\nu}{}^{a}$. 
The local interactions among 
$\widehat{B}_{\mu\nu}{}^{a}$, $\widehat{C}_{\mu\nu}{}^{a}$ and the local 
Yang-Mills fields $A_{\mu}{}^{a}$ are determined by the non-linear 
realization method developed for the loop gauge group. The 
central charge $k$ does not appear in these interactions.

Let us now derive the action $S_{{\rm R}}^{U(1)}$ written in 
terms of the local fields. As was done in Sec. 3, substituting (5.44) into 
(5.50), we obtain 
${\cal L}^{U(1)}$ in terms of the local fields. Next, inserting the 
Lagrangian into (5.48), and carrying out the integrations over 
$x^{\mu n} \,(n \neq 0)$ for (5.48), we obtain the action 
$S_{{\rm R}}^{U(1)}$ written in terms of the local fields. The concrete form 
of $S_{{\rm R}}^{U(1)}$ is given by
\begin{eqnarray}
& & S_{{\rm R}}^{U(1)} = \int d^{D}x \Biggl[ \Biggr.
 -\frac{1}{4} \Biggl\{ \Biggr. 
 R_{\mu\gamma\xi\eta}{}^{\nu\zeta\kappa\lambda}
 \partial^{\,\mu}\widetilde{B}^{\gamma\xi\eta}(x) 
 \partial_{\nu}\widetilde{B}_{\zeta\kappa\lambda}(x)
\nonumber \\
& & \quad 
 + a_{1}R_{\xi\eta}{}^{\kappa\lambda}
 \left( \partial_{\kappa}\widetilde{B}_{[\mu\nu]\lambda}(x) 
 + \partial_{\,[\mu}\widetilde{B}_{\nu]\kappa\lambda}(x) \right) 
\nonumber \\
& & \quad \qquad 
 \times \left( \partial^{\,\xi}\widetilde{B}^{\mu\nu\eta}(x) 
 + \partial^{\,\mu}\widetilde{B}^{\nu\xi\eta}(x) \right) 
\nonumber \\
& & \quad 
 + 4a_{2}R_{\xi\eta\gamma}{}^{\mu\kappa\lambda}
 \partial^{\,\gamma}\widetilde{B}^{\,\nu\xi\eta}(x) 
\nonumber \\
& & \quad \qquad 
 \times \left( 4\partial_{\kappa}\widetilde{B}_{\,[\mu\nu]\lambda}(x) 
 + 4\partial_{\,[\mu}\widetilde{B}_{\nu]\kappa\lambda}(x) 
 + \partial_{\mu}\widetilde{B}_{\nu\kappa\lambda}(x) \right) 
\nonumber \\
& & \quad 
 + a_{3}R_{\mu\gamma\xi\eta}{}^{\nu\zeta\kappa\lambda}
 \partial^{\,\mu}\widetilde{C}^{\gamma\xi\eta}(x) 
 \partial_{\nu}\widetilde{C}_{\zeta\kappa\lambda}(x)
\nonumber \\
& & \quad 
 + a_{2}R_{\xi\eta\gamma}{}^{\kappa\lambda\zeta}
 \left( 2\partial_{\mu}\widetilde{C}_{\kappa\lambda\zeta}(x) 
   -3\partial_{\kappa}\widetilde{C}_{\mu\lambda\zeta}(x) \right) 
\nonumber \\
& & \quad \qquad 
 \times \left( 2\partial^{\,\mu}\widetilde{C}^{\,\xi\eta\gamma}(x) 
 - 3\partial^{\,\xi}\widetilde{C}^{\mu\eta\gamma}(x) \right) 
\nonumber \\
& & \quad 
 - 2a_{2}R_{\xi\eta\gamma}{}^{\mu\kappa\lambda}
 \left( 2\partial^{\,\nu}\widetilde{C}^{\,\xi\eta\gamma}(x)
  - 3\partial^{\,\xi}\widetilde{C}^{\,\nu\eta\gamma}(x) \right) 
\nonumber \\
& & \quad \qquad 
 \times \left( 2\partial_{\kappa}\widetilde{B}_{\,[\mu\nu]\lambda}(x) 
 + 2\partial_{\,[\mu}\widetilde{B}_{\nu]\kappa\lambda}(x) 
 + \partial_{\mu}\widetilde{B}_{\nu\kappa\lambda}(x) \right) 
\nonumber \\
& & \quad 
 -2a_{3}R_{\mu\gamma\xi\eta}{}^{\nu\zeta\kappa\lambda}
 \partial_{\nu}\widetilde{B}_{\zeta\kappa\lambda}(x)
 \partial^{\,\mu}\widetilde{C}^{\gamma\xi\eta}(x) 
\nonumber \\
& & \quad 
 + 2\tilde{k}_{1}a_{1}R_{\xi\eta}{}^{\kappa\lambda}
 \widehat{B}^{\,\xi\mu}{}_{a}(x)F^{\nu\eta,a}(x) 
 \left( \partial_{\kappa}\widetilde{B}_{[\mu\nu]\lambda}(x) 
  + \partial_{\,[\mu}\widetilde{B}_{\nu]\kappa\lambda}(x) \right)
\nonumber \\
& & \quad 
 - 4\tilde{k}_{1}a_{2}R_{\xi\eta\gamma}{}^{\mu\kappa\lambda}
 \widehat{C}^{\,\xi\eta}{}_{a}(x)F^{\nu\gamma, a}(x) 
\nonumber \\
& & \quad \qquad 
 \times \left( 2\partial_{\kappa}\widetilde{B}_{\,[\mu\nu]\lambda}(x) 
 + 2\partial_{\,[\mu}\widetilde{B}_{\nu]\kappa\lambda}(x) 
 + \partial_{\mu}\widetilde{B}_{\nu\kappa\lambda}(x) \right) 
\nonumber \\
& & \quad 
 + 2\tilde{k}_{1}a_{2}R_{\xi\eta\gamma}{}^{\kappa\lambda\zeta}
 \widehat{C}^{\,\xi\eta}{}_{a}(x)F^{\mu\gamma, a}(x) 
\nonumber \\
& & \quad \qquad 
 \times \left( 2\partial_{\mu}\widetilde{C}_{\kappa\lambda\zeta}(x)
 - 3\partial_{\kappa}\widetilde{C}_{\mu\lambda\zeta}(x) \right) 
\nonumber \\
& & \quad 
 + \tilde{k}_{1}{}^{2}a_{1}R_{\xi\gamma}{}^{\kappa\lambda}
 \widehat{B}_{\kappa[\mu}{}^{a}(x)F_{\nu]\lambda, a}(x)
 \widehat{B}^{\,\xi\mu}{}_{b}(x)F^{\nu\gamma, b}(x) 
\nonumber \\
& & \quad 
 + 4\tilde{k}_{1}{}^{2}a_{2}R_{\xi\eta\gamma}{}^{\kappa\lambda\zeta}
 \widehat{C}_{\kappa\lambda}{}^{a}(x)F_{\mu\zeta, a}(x)
 \widehat{C}^{\,\xi\eta, b}(x)F^{\mu\gamma}{}_{b}(x) 
 \; \Biggl. \Biggr\}
\nonumber \\
& & + \frac{1}{2}m_{q1}{}^{2} \Biggl\{ \Biggr.
 \widetilde{B}_{\mu\nu\kappa}(x)
 \widetilde{B}^{\,\mu\nu\kappa}(x)
 + b_{1}\widetilde{B}_{\mu\nu\kappa}(x)
 \widetilde{B}^{\,\nu\mu\kappa}(x) 
\nonumber \\
& & \quad 
 + b_{2}\widetilde{B}_{\mu\kappa}{}^{\kappa}(x)
 \widetilde{B}^{\,\mu\lambda}{}_{\lambda}(x)
 + b_{3}\widetilde{B}^{\,\kappa}{}_{\kappa\mu}(x)
 \widetilde{B}^{\,\mu\lambda}{}_{\lambda}(x) 
\nonumber \\
& & \quad 
 + b_{4}\widetilde{B}^{\,\kappa}{}_{\kappa\mu}(x)
 \widetilde{B}_{\lambda}{}^{\lambda\mu}(x) 
 + b_{5}\widetilde{C}_{\mu\nu\kappa}(x)
 \widetilde{C}^{\,\mu\nu\kappa}(x) 
\nonumber \\
& & \quad 
+ b_{6}\widetilde{C}^{\,\kappa}{}_{\kappa\mu}(x)
 \widetilde{C}^{\,\mu\lambda}{}_{\lambda}(x) 
+ b_{7}\widetilde{B}_{\mu\nu\kappa}(x)
 \widetilde{C}^{\,\mu\nu\kappa}(x) 
\nonumber \\
& & \quad 
 + b_{8}\widetilde{B}_{\mu\kappa}{}^{\kappa}(x)
 \widetilde{C}^{\,\mu\kappa}{}_{\kappa}(x)
 + b_{9}\widetilde{B}^{\,\kappa}{}_{\kappa\mu}(x)
 \widetilde{C}^{\,\mu\lambda}{}_{\lambda}(x) 
 \Biggl. \Biggr\} \Biggl. \Biggr]  \;,
\end{eqnarray}
%
with
{
\setcounter{enumi}{\value{equation}}
\addtocounter{enumi}{1}
\setcounter{equation}{0}
\renewcommand{\theequation}{\thesection.\theenumi\alph{equation}}
\begin{eqnarray}
& & R_{\mu\nu}{}^{\kappa\lambda} 
= \delta_{(\mu}{}^{\kappa}\delta_{\nu)}{}^{\lambda}
+ \eta_{\mu\nu}\eta^{\kappa\lambda} \;, \\
& & R_{\mu\nu\xi}{}^{\kappa\lambda\gamma} 
= \delta_{(\mu}{}^{\kappa}\delta_{\nu}{}^{\lambda}\delta_{\xi)}{}^{\gamma}
+ \frac{1}{4}\eta_{(\mu\nu}\eta^{(\kappa\lambda}\delta_{\xi)}{}^{\gamma)} \;,\\
& & R_{\mu\nu\xi\eta}{}^{\kappa\lambda\gamma\zeta} 
= \delta_{(\mu}{}^{\kappa}\delta_{\nu}{}^{\lambda}
 \delta_{\xi}{}^{\gamma}\delta_{\eta)}{}^{\zeta}
+ \frac{1}{8}\eta_{(\mu\nu}\eta^{(\kappa\lambda}
 \delta_{\xi}{}^{\gamma}\delta_{\eta)}{}^{\zeta)} \nonumber \\
& & \quad + \frac{1}{64}\eta_{(\mu\nu}\eta_{\xi\eta)}
 \eta^{(\kappa\lambda}\eta^{\gamma\zeta)} \;.
\end{eqnarray}
%
Here $a_{i}\; (i=1,2,3)$ and $b_{i} \;(i=1,2,\ldots,9)$ are constants given by 
\setcounter{equation}{\value{enumi}}
}
\begin{eqnarray*}
& & a_{1}=(D+4)(D+6)\;, \quad 
    a_{2}=\frac{D+6}{16} \;, \quad
    a_{3}=\frac{1}{16} \;, \\
& & b_{1}=\frac{1}{2}J 
 \{ 4K(2D-1)(D+3)\delta(0) 
 - 16(D-1)(D^{2}+4D+1)\delta''(0) \} \;, \\
& & b_{2}=-\frac{1}{2}J
 \{ K(3D^{2}+22D-41)\delta(0) 
 + 4(D-1)(3D+11)\delta''(0) \} \;, \\
& & b_{3}=J
 \{ 4K(D+1)^{2}\delta(0)+16(D-1)(D+3)\delta''(0) \} \;, \\
& & b_{4}=-J
 \{ 12K(D+1)\delta(0)+16(D-1)\delta''(0) \} \;, \\
& & b_{5}=-\frac{1}{2}J
 \{ 27K(D+1)^{2}\delta(0)+12(D-1)(D+21)\delta''(0) \} \;, \\
& & b_{6}=\frac{1}{4}J
 \{ 9K(D^{2}-18D-7)\delta(0) 
 - 36(D-1)(D-13)\delta''(0) \} \;, \\
& & b_{7}=6J
 \{ 3K(D+1)^{2}\delta(0)-2(D-1)(3D+7)\delta''(0) \} \;, \\
& & b_{8}=-3J
 \{ K(D^{2}+10D-3)\delta(0)+2(D-1)^{2}\delta''(0) \} \;, \\
& & b_{9}=3J
 \{ 2K(D+1)(D-5)\delta(0)-4(D-1)(3D+7)\delta''(0) \} \;,\end{eqnarray*}
with 
\begin{eqnarray*}
& & K \equiv -\frac{1}{V_{{\rm R}}}\int \{dx \}
 \int^{2\pi}_{0} \frac{d\sigma}{2\pi} Q'_{\mu}(\sigma)Q'^{\mu}(\sigma)
 \exp \left( -\frac{L}{l^{2}} \right) (>0) \;, \\
 \nonumber \\
& & J^{-1} \equiv 2K(4D^{3}+19D^{2}-20D-15)\delta(0) 
 - 8(D-1)(2D^{2}+9D+5)\delta''(0) \;.
\end{eqnarray*}
In deriving the action (5.51), we have set the free parameters $k_{u}$, 
$q_{1}$, and $\tilde{q}_{1}$ so as to satisfy the normalization conditions 
{
\setcounter{equation}{\value{enumi}}
\setcounter{enumi}{\value{equation}}
\addtocounter{enumi}{1}
\setcounter{equation}{0}
\renewcommand{\theequation}{\thesection.\theenumi\alph{equation}}
\begin{eqnarray}
& & \frac{k_{u}q_{1}{}^{2}l^{2}}{4(D+2)(D+4)(D+6)}
 \delta(0)^{2} = -1 \;, \\
& & \frac{k_{u}\tilde{q}_{1}{}^{2}}{D(D-1)(D+2)(D+4)} 
\times \{ 2K(4D^{3}+19D^{2}-20D-15)\delta(0) 
\nonumber \\ 
& & \qquad \quad 
 - 8(D-1)(2D^{2}+9D+5)\delta''(0) \} = -1 \;. 
\end{eqnarray}
%
The action (5.51) describes the massive tensor fields theory for 
$\widetilde{B}_{\mu\nu\lambda}$ and $\widetilde{C}_{\mu\nu\lambda}$ 
without spoiling the gauge invariance. This property is also possessed by 
St\"uckelberg formalism. 
Reflecting the non-abelian gauge theory, however, the action (5.51) 
includes non-trivial couplings via the {\it BF}-like terms 
$\widehat{B}_{\mu\nu}{}^{a}F_{\lambda\kappa, a}$ and 
$\widehat{C}_{\mu\nu}{}^{a}F_{\lambda\kappa, a}$. 
It is obvious that these couplings are due to 
the central extension of the gauge group. 
Indeed, all the interaction terms occurring in (5.51) include 
the central charge $k$. 
By setting $k=0$, we find that the gauge invariant tensor fields 
$\widetilde{B}_{\mu\nu\lambda}$ and $\widetilde{C}_{\mu\nu\lambda}$ reduce 
to the components of $\widetilde{A}_{\mu\nu\lambda}$, and all 
the interaction terms occurring in (5.51) vanish. 
Hence, (5.51) becomes the action for the massive 
tensor field $\widetilde{A}_{\mu\nu\lambda}$ without interactions. 
The gauge invariance still holds, because 
$\widetilde{A}_{\mu\nu\lambda}$ is invariant under the 
transformation rules of (5.15) and (5.21) with $k=0$. 
Therefore, (5.51) results in the action 
of the St\"uckelberg formalism for the abelian tensor 
field of third rank $\widetilde{A}_{\mu\nu\lambda}$. \cite{LSU(1)T} 
Consequently, we can regard (5.51) as 
the action of the ``generalized'' St\"uckelberg formalism for the tensor 
fields of third rank $\widetilde{B}_{\mu\nu\lambda}$ and 
$\widetilde{C}_{\mu\nu\lambda}$ in a broad sense.

We next comment on the interactions in (5.51). 
Although the types of 
interactions in (5.51) are somewhat complicated, we can find 
some features of the interactions. 
First, the abelian tensor fields $\widetilde{B}_{\mu\nu\lambda}$ and 
$\widetilde{C}_{\mu\nu\lambda}$ couple with the non-abelian tensor fields 
$\widehat{B}_{\mu\nu}{}^{a}$ and $\widehat{C}_{\mu\nu}{}^{a}$ and the local 
Yang-Mills fields $A_{\mu}{}^{a}$ via the {\it BF}-like terms. 
(Here, $\widetilde{C}_{\mu\nu\lambda}$ does 
not couple with $\widehat{B}_{\mu\nu}{}^{a}F_{\lambda\kappa, a}$.) 
Second, the second power of the {\it BF}-like terms give couplings among 
the non-abelian fields $\widehat{B}_{\mu\nu}{}^{a}$, 
$\widehat{C}_{\mu\nu}{}^{a}$ and $A_{\mu}{}^{a}$  
that are obviously different from the 
minimal interactions resulting from the covariant derivative. \cite{LSYM} 
We would like to emphasize that these features are analogous to those 
of the couplings in the action (3.19). Instead of the Chern-Simons term, 
the {\it BF}-like terms contribute to the non-trivial couplings in the 
action (5.51). We may regard the couplings between the abelian tensor fields 
and the {\it BF}-like terms as a kind of generalization of the 
Chapline-Manton coupling. 

\setcounter{equation}{\value{enumi}}
}


\section{Conclusion and discussion}
\setcounter{equation}{0}

In this paper, we have considered the EYMT in loop space whose gauge group is 
the affine Lie group. In the EYMT, central extension of the gauge 
group leads to a coupling between the Yang-Mills 
fields ${\cal A}_{\mu\sigma}^{Y}$ and the {\it U}(1) gauge 
field ${\cal A}_{\mu\sigma}^{U}$. The coupling is different from the 
minimal coupling and the coupling via the Pauli terms existing in the 
Standard Model. \cite{WS} 
The coupling yields non-trivial couplings between non-abelian local fields 
included in the Yang-Mills fields ${\cal A}_{\mu\sigma}^{Y}$ and an abelian 
local field included in the {\it U}(1) gauge field, ${\cal A}_{\mu\sigma}^{U}$. 
The Chapline-Manton coupling, which was originally introduced in order 
to combine a supergravity and a super Yang-Mills system, can be 
systematically derived within the framework of the Yang-Mills theory. 
In the supergravity theory, the Chapline-Manton coupling is derived 
using the local supersymmetry. \cite{ACM} It is interesting to study the 
relation of the central extension of the gauge group and local supersymmetry.

By using the formalism of the non-linear realization developed for 
the affine Lie gauge group, furthermore, we can derive the 
``generalized'' Chapline-Manton coupling for higher-rank tensor fields. 
This coupling is given by the couplings among the local 
Yang-Mills fields $A_{\mu}{}^{a}$, the non-abelian tensor fields of 
second rank $\widehat{B}_{\mu\nu}{}^{a}$ and $\widehat{C}_{\mu\nu}{}^{a}$, and 
the abelian tensor fields of third rank $\widetilde{B}_{\mu\nu\lambda}$ 
and $\widetilde{C}_{\mu\nu\lambda}$ via the {\it BF}-like terms. 
In the (bosonic) string theory, an abelian antisymmetric tensor field of 
second rank appears as massless excited states, while an abelian tensor field 
of third rank having the same symmetric property as 
$\widetilde{A}_{\mu\nu\lambda}$ \,($\widetilde{A}_{\mu\nu\lambda}=
\widetilde{A}_{\mu\lambda\nu}$) 
appears as massive excited states. \cite{MST} 
The Chapline-Manton coupling is realized in the type I supergravity theory. 
This theory is obtained as the low energy effective theory of the type I (or 
heterotic) superstring theory. The generalized Chapline-Manton 
coupling including the abelian tensor field of third rank might 
be realized in a massive mode in string theory.

If an abelian tensor field couples with the {\it BF} term, then it must have 
the totally antisymmetric property, because {\it BF} terms have this 
property. However, both the abelian tensor fields of 
third rank $\widetilde{B}_{\mu\nu\lambda}$ and $\widetilde{C}_{\mu\nu\lambda}$ 
have some specific symmetric properties. For this reason, these abelian tensor 
fields 
cannot couple with the {\it BF} term. Such a difficulty as this might 
be settled by considering the EYMT in {\it closed p-manifold space} 
$\Omega^{p} M^{D}$, which is the configuration space for closed 
{\it p}-branes.  \cite{PLS} \cite{PMS} 
A {\it U}(1) gauge field ${\cal A}^{U(0)}_{\mu\sigma}[x]$ on 
$\Omega^{p} M^{D}$ consisting of an abelian local tensor fields is given by 
\begin{eqnarray}
{\cal A}^{U(0)}_{\mu\vec{\sigma}}[x]
 = \frac{q_{0}}{p!}\Sigma^{\nu_{1}\nu_{2} \cdots \nu_{p}}(\vec{\sigma})\,
 B_{\mu\nu_{1}\nu_{2} \cdots \nu_{p}}(x(\vec{\sigma})) \;, 
\end{eqnarray}
%
where $B_{\mu\nu_{1}\nu_{2} \cdots \nu_{p}}(x)$ is an (abelian) 
totally antisymmetric tensor field of rank $(p+1)$ on $M^{D}$ and 
$\Sigma^{\nu_{1}\nu_{2} \cdots \nu_{p}}
(\vec{\sigma}) \equiv x'_{1}{}^{[\nu_{1}}(\vec{\sigma})x'_{2}{}^{\nu_{2}}
(\vec{\sigma}) \cdots x'_{p}{}^{\nu_{p}]}(\vec{\sigma})$. 
[Here, $\vec{\sigma} = (\sigma_{1}, \sigma_{2}, \ldots , \sigma_{p})$ 
represents the parameters describing a closed {\it p}-brane and 
$x'^{\mu}_{n}(\vec{\sigma}) \equiv 
\partial x^{\mu}(\vec{\sigma})/\partial \sigma_{n} \;\; 
( 1 \leq n \leq p )$.]  We can indeed obtain the local field theory of 
$B_{\mu\nu_{1}\nu_{2} \cdots \nu_{p}}(x)$ from the {\it U}(1) gauge theory in 
closed {\it p}-manifold space. \cite{PMS} 
In order to carry out a similar extension to the Yang-Mills theory, 
we have to find a suitable gauge group other than the affine Lie gauge group. 
It is conceivable that the suitable gauge group for the Yang-Mills theory in 
closed {\it p}-manifold space is the Mickelson-Faddeev group (and its 
generalization to higher dimensions). \cite{ALG} \cite{MF} \cite{MF2} 
The commutation relations of the generators of the (generalized) 
Mickelson-Faddeev group are given by 
\begin{eqnarray}
& & [\,T_{a}(\vec{\rho}), T_{b}(\vec{\sigma})\,] 
= if_{ab}{}^{c}T_{c}(\vec{\sigma})\delta^{p}(\vec{\rho}-\vec{\sigma}) 
\nonumber \\
& & \qquad \qquad + \, k  \epsilon^{j_{1}j_{2} \cdots j_{p}}
 \partial_{j_{1}}T_{(ab)}{}_{j_{2} \cdots j_{p-1}}(\vec{\sigma})
 \partial_{j_{p}}\delta^{(p)}(\vec{\rho}-\vec{\sigma}) \;.
\end{eqnarray}
%
Setting $p=1$, we find that (6.2) coincides with (2.1). 
The commutation relations (6.2) are a natural extension of (2.1).
We hope to discuss this subject in the future.


\section*{Acknowledgements}

I am grateful to S. Deguchi for valuable discussions and helpful suggestions. 
I would also like to thank A. Sugamoto for his careful reading of the 
manuscript. Thanks are also due to S. L. Shatashvili for information 
regarding Ref. 23) 

\clearpage


\end{document}